\documentclass[fleqn,usenatbib]{mnras}

\usepackage[T1]{fontenc}
\usepackage{graphicx}	
\usepackage{amsmath}	
\usepackage{amssymb}	
\usepackage{url}        
\usepackage{xcolor}     
\usepackage{subcaption} 
\usepackage{newtxtext,newtxmath}

\DeclareRobustCommand{\VAN}[3]{#2}
\let\VANthebibliography\thebibliography
\def\thebibliography{\DeclareRobustCommand{\VAN}[3]{##3}\VANthebibliography}

\newcommand{\red}[1]{\textcolor{red}{#1}}

\newcommand{\gcm}{g\,cm$^{-3}$ }	
\newcommand{\ergg}{erg g$^{-1}$ } 
\newcommand{\dynecm}{dyne cm$^{-2}$ } 
\newcommand{\gs}{g s$^{-1}$ } 

\title[White dwarf planetary debris from asteroid belts]{White dwarf planetary debris dependence on physical structure distributions within asteroid belts}

\author[McDonald \& Veras]{
Catriona H. McDonald,$^{1, 2}$\thanks{E-mail: catriona.mcdonald@warwick.ac.uk}
Dimitri Veras,$^{1, 2}$\thanks{STFC Ernest Rutherford Fellow}
\\
$^{1}$Centre for Exoplanets and Habitability, University of Warwick, Coventry CV4 7AL, UK\\
$^{2}$Department of Physics, University of Warwick, Coventry CV4 7AL, UK \\
}

\date{Accepted XXX. Received YYY; in original form ZZZ}

\pubyear{2021}

\begin{document}
\label{firstpage}
\pagerange{\pageref{firstpage}--\pageref{lastpage}}
\maketitle

\begin{abstract}
White dwarfs which exhibit transit signatures of planetary debris and accreted planetary material provide exceptional opportunities to probe the material composition and dynamical structure of planetary systems. 
Although previous theoretical work investigating the role of minor body disruption around white dwarfs has focussed on spherical bodies, Solar System asteroids can be more accurately modelled as triaxial ellipsoids. 
Here we present an analytical framework to identify the type of disruption (tidal fragmentation, total sublimation or direct impact) experienced by triaxial asteroids approaching white dwarfs on extremely eccentric ($e \sim 1$) orbits. 
This framework is then used to identify the outcomes for simplified Main belt analogues of 100 bodies across five different white dwarf temperatures. 
We also present an empirical relationship between cooling age and effective temperature for both DA and DB white dwarfs to identify the age of the white dwarfs considered here.
We find that using a purely spherical shape model can underestimate the physical size and radial distance at which an asteroid is subjected to complete sublimation, and these differences increase with greater elongation of the body.
Contrastingly, fragmentation always occurs in the largest semi-axis of a body and so can be modelled by a sphere of that radius. 
Both fragmentation and sublimation are greatly affected by the body's material composition, and hence by the composition of their progenitor asteroid belts.
The white dwarf temperature, and hence cooling age, can affect the expected debris distribution: higher temperatures sublimate large elongated asteroids, and cooler temperatures accommodate more direct impacts. 
\end{abstract}




\section{Introduction} \label{sec:intro}
White dwarfs provide a unique opportunity to investigate the composition of exoplanetary bodies. 
The extreme surface gravities of white dwarfs cause elements heavier than hydrogen or helium to rapidly sink and not be visible in spectra \citep{Paquette1986, Wyatt2014}.
However, observations indicate that between a quarter and half of all white dwarfs have evidence of metals in their atmospheres \citep{Zuckerman2010, Koester2014}, with the most commonly visible elements being closely aligned with the composition of the Solar System terrestrial planets \citep{Jura2014, Hollands2018, Doyle2019}.
The consistent visibility of these metals suggest ongoing accretion of planetary material.

Recent observations have found evidence of planetary bodies which could lead to this accretion. 
\cite{Vanderburg2015} identifies at least one, but most likely at least six, rocky planetesimals with densities $>2$~\gcm, actively disrupting around the white dwarf WD~1145+017.
These planetesimals orbit on short periods ($\sim 4.5-4.9$h) near the Roche limit of the star, causing them to be frequently releasing material which forms a dust cloud, observed in asymmetric transit curves of up to $60$ per cent in depth\footnote{See \url{http://www.brucegary.net/1145/} for detailed observations of the debris around WD~1145+017 between 2015-19.} \citep[See also][]{Gansicke2016, Rappaport2016, Zhou2016, Croll2017, Gary2017, Izquierdo2018, Vanderburg2018}.

\cite{Vanderbosch2020} report the observation of a planetesimal on a highly eccentric ($e > 0.97$) orbit around the white dwarf ZTF~J0139+5245 producing transit depths of up to $45$ per cent.
The observations indicate that the object is in an early stage of disruption.
However, it's large $110$-day period could be indicative of an orbital pericentre outside of the star's Roche limit. 
The planetesimal's disruption could therefore originate from an alternative mechanism to the canonical Roche limit disruption, which doesn't involve tidal forces from the star. \cite{Veras2020a} show that chaotic exchange of orbital and spin angular momentum can lead to an ellipsoidal planetesimal achieving a spin rate higher than the cohesionless spin barrier \citep[see fig. 1 of][]{Warner2009} and disrupting.

Most recently \cite{Vanderbosch2021} reported a third transiting minor body around the white dwarf ZTF~J0328-1219, exhibiting two significant periods at $9.937$ hours and $11.2$ hours.
The shorter period is roughly twice that of the debris orbiting WD~1145+017 and much less than that around ZTF~J0139+5245.
These transits show much shallower depths of $~10$ per cent and exhibit variability across the entire phase of the orbit, which suggests this object is in a different stage of disruption compared to the two previously discovered objects.

These three minor bodies actively disrupting in different orbital configurations raise questions about the circumstances that lead to planestesimal disruption around white dwarfs, and the object around ZTF~J0139+5245 showcases the importance of adopting aspherical asteroid models.
Further, \cite{Guidry2021} reported an additional two candidates exhibiting variation on long timescales similar to ZTF~J0139+5245 and two more with shorter variations akin to WD~1145+017, highlighting an urgent need to increase our understanding of the disruption process.

During the giant branch phases of a star's evolution, 0.1-10 km bodies within $\sim 7$ au of the star will be broken down to their strongest components by the YORP effect \citep{Veras2014b}.
It is likely that bodies at larger distances or with high internal strengths \citep{Veras2020e} will remain intact despite both luminosity variations and the dynamical instability of the remnant planetary system after the giant branch mass loss phases \citep{Debes2002, Veras2015a, Veras2016b, Mustill2018, Maldonado2020c, Maldonado2020b, Maldonado2020a}.
Minor bodies can then be vulnerable to perturbations from major bodies like more distant analogues of the gas giants recently discovered by \cite{Gansicke2019} and \cite{Vanderburg2020} and approach the white dwarf on eccentric orbits.

A further $1-3$ per cent of white dwarfs display infrared excesses indicative of dusty debris discs \citep{RebassaMansergas2019}, with $0.04-0.1$ per cent also having an observed gaseous component \citep{Gansicke2006, Manser2020}.
Although it should be noted that discs, or narrower rings of debris, should exist around most polluted white dwarfs, with an estimated 90 per cent of all such discs being currently unobservable \citep{Rocchetto2015}. 
It is generally thought that the perturbed minor bodies come within the star's Roche limit and are tidally disrupted, forming the observed debris discs \citep{Debes2012, Veras2014a, Malamud2020a}.

Numerical simulations of debris disc formation suggest a minimum disc mass of $\sim 10^{23}$ g to agree with observations \citep{Kenyon2017, VanLieshout2018, Farihi2018}, which is comparable to the mass of the largest Main belt asteroid Ceres and the theoretically constrained mass of the asteroid disrupting around WD~1145+017 \citep{Rappaport2016, Gurri2017}.
The first white dwarf observed with an infrared excess caused by a dusty debris disc, G29-38 \citep{Zuckerman1987}, is estimated to have accreted $\sim 4 \times 10^{24}$~g of material, about the total mass of the asteroids in the Solar System \citep{Jura2003}. 

The observational mass and chemical abundance constraints, alongside the abundance and dynamical availability of asteroids, have led to these minor bodies being the preferred cause of white dwarf pollution. 
The observed accreted material is largely terrestrial in composition, which suggests the polluting bodies will have formed within the snow line of their planetary system \citep{Martin2020}.
A typical $0.6 M_\odot$ white dwarf would have a $1.39 \pm 0.44 M_\odot$ main sequence progenitor \citep[see eq. 4 of][]{Cummings2018}, with a water ice line at $\sim 2$ au \citep{Adams1986, Kenyon1987, Chiang1997, Kennedy2008}. 
Although, it should be noted that a small number of white dwarf systems show evidence for the accretion of more icy, Kuiper belt like bodies \citep{Farihi2013, Raddi2015, GentileFusillo2017, Xu2017, Hoskin2020}.
Further, dynamical mixing between the terrestrial Main belt and volatile Kuiper belt in the Solar System should populate the Main belt region with both rocky and icy bodies.
Thus, pollutant asteroids could have either a rocky terrestrial composition or a volatile rich icy composition.

Solar System asteroids have been well observed and studied, revealing a wide range of shapes, sizes and characteristics \citep{Warner2009, Durech2018}.
Observed orbital and physical parameters of these bodies have been successfully reproduced using ellipsoidal rather than spherical models \citep{Carbognani2012, Dobrovolskis2019}.
Using this knowledge, this paper aims to expand on the work presented in \cite{Brown2017} (hereafter BVG17), which investigates the destruction of quasi-spherical bodies approaching a white dwarf on a parabolic trajectory.
BVG17 formed a basic first step in understanding the role of asteroids in white dwarf pollution, but their results may not be suitably accurate for comparison to observations, or with other theory \citep[e.g.][]{Wyatt2014}.
The following work (i) considers the effect of imposing asphericity on the asteroid treatment presented in BVG17, and (ii) applies the formalism to Main belt analogue reservoirs of white dwarf pollutants.

In Section~\ref{sec:properties} we introduce the properties of the white dwarfs considered in this paper (Section~\ref{subsec:wd}), the shape and material properties of asteroids (Section~\ref{subsec:asteroid_props}) and the properties of asteroid Main belt analogues (Section~\ref{subsec:belt_props}).
In Section~\ref{sec:approach} we introduce the conditions for the three different disruption regimes (i) \textit{sublimation}, (ii) \textit{fragmentation} and (iii) \textit{impact} and the analytical formalism used to identify which form of disruption will befall a particular asteroid.
We then apply this formalism to a Main belt analogue of 100 asteroids in Section~\ref{sec:main_belt} and further investigate the role of triaxiality in the disruption regime an asteroid befalls. 
Section~\ref{sec:further} briefly considers what happens to the asteroidal material after the initial disruption process identified in Section~\ref{sec:main_belt}. We conclude in Section~\ref{sec:conclusions}.

\section{Properties} \label{sec:properties}

We begin by characterizing the physical and orbital properties separately of white dwarfs (Section \ref{subsec:wd}), asteroids (Section \ref{subsec:asteroid_props}) and asteroid belts (Section \ref{subsec:belt_props}).

\subsection{White Dwarfs} \label{subsec:wd}
Although the majority of stars in the Milky Way will become white dwarfs \citep{Koester2013}, white dwarf masses are restricted to a very small range, typically between $0.4-0.8 M_\odot$, although the most massive white dwarfs can have $M_{\rm WD} \approx 1.4 M_\odot$. In the following, a white dwarf mass of $0.6M_\odot$ is used, which corresponds to the peak of the white dwarf mass distribution \citep{Althaus2010, Kleinman2013, Tremblay2016, McCleery2020}.

\subsubsection{Radii} \label{subsubsec:wdmass_rad}

White dwarf radii are closely related to their masses through a mass-radius relationship \citep{Hamada1961}.
BVG17 utilised a basic mass-radius relationship, which exploited the fact the relation is approximately independent of temperature as follows
\begin{equation}
    R_\text{WD} = \gamma R_\odot \left( \frac{M_\text{WD}}{M_\odot} \right)^{-1/3},
    \label{eq:mass-radius_BVG}
\end{equation}
with $\gamma \simeq 10^{-2}$. 

Here, we use a more precise form of the mass-radius relationship \citep[eqs 27-28 of][]{Nauenberg1972}
\begin{equation}
\frac{R_{\rm WD}}{R_{\odot}}
\approx
0.0127 
\left( \frac{M_{\rm WD}}{M_{\odot}} \right)^{-1/3}
\sqrt{
1 - 0.607
\left( \frac{M_{\rm WD}}{M_{\odot}} \right)^{4/3}
}
,
\label{eq:mass-radius_Nau}
\end{equation}

\noindent by assuming a mean molecular weight per electron of $2$ from \cite{Hamada1961}.

\subsubsection{Temperature and cooling age} \label{subsubsec:wdtemp_cool}

During the white dwarf phase, nuclear burning no longer proceeds, so although white dwarfs can start off at very high temperatures, they then monotonically cool for the rest of their lifetimes.
How long a star has been in the white dwarf phase is known as the cooling age, $\tau_{{\text{WD}}_{\text{cool}}}$, and is a function of the temperature and mass of the white dwarf.

Here we derive an empirical relation between a white dwarf's cooling age and effective temperature $T_{\rm eff}$. This algebraic relation may potentially be useful for and facilitate future studies. We present both a long-form relation for $\tau_{{\rm WD}_{\rm cool}}$ as a function of both $M_\text{WD}$ and $T_{\rm eff}$ and a more compact relation (useful for analytical manipulations) as a function of just $T_{\rm eff}$ for a $0.6 M_\odot$ white dwarf.

Our formulae attempt to match the cooling models of {\it both} DA and DB white dwarfs (with the same relation) from \cite{Fontaine2001}. These models can be downloaded
\footnote{\url{http://www.astro.umontreal.ca/~bergeron/CoolingModels}} and should be used for higher precision work; here we seek just a rough estimate with an analytic formula. Our long-form relation is. 

\begin{figure}
	\includegraphics[width=\columnwidth]{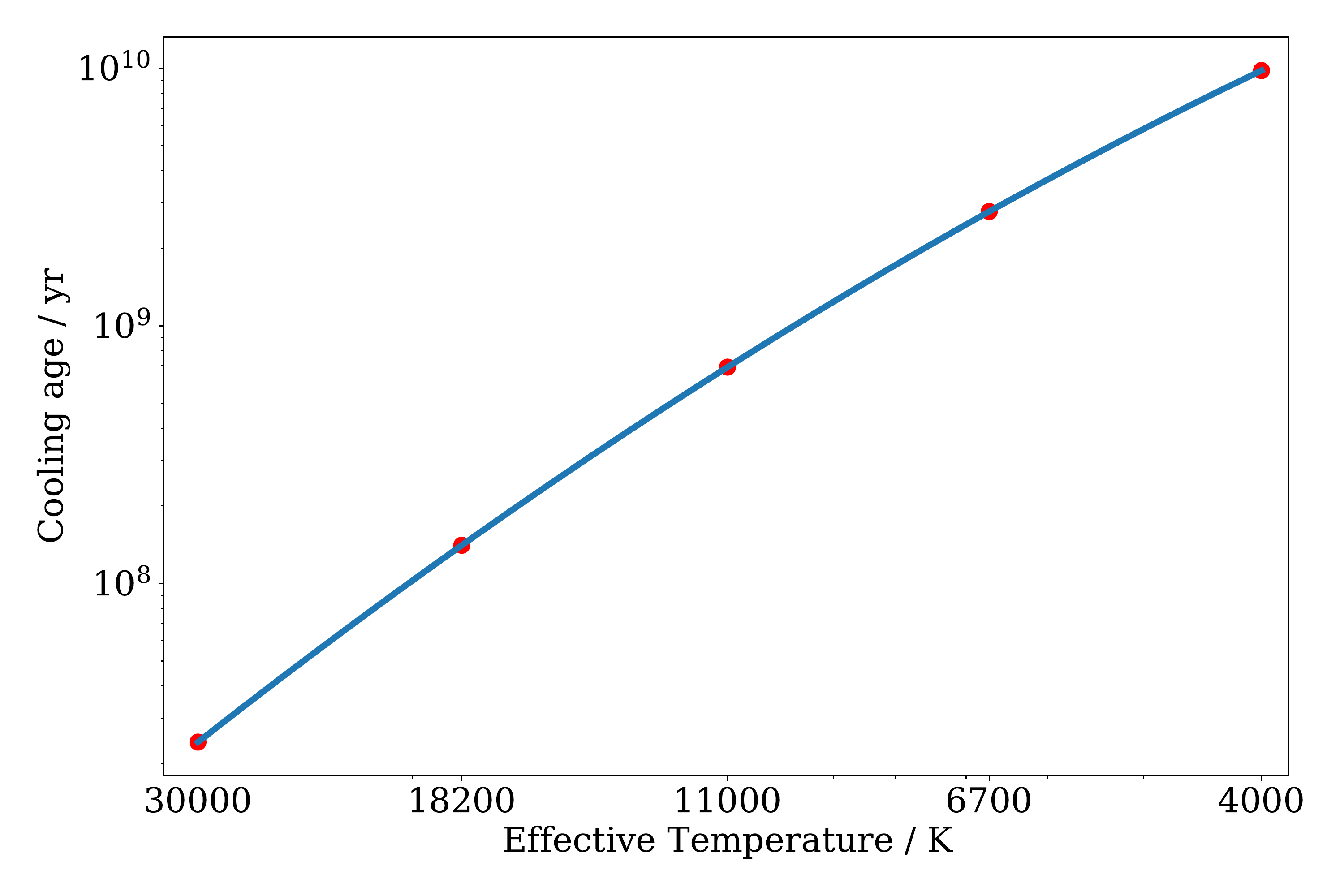}
    \caption{Empirical formula relating white dwarf cooling age to effective temperature for a $0.6M_\odot$ white dwarf.
    The red circles indicate the five equally log-spaced temperatures chosen for analysis in this work. 
    From largest temperature to smallest the cooling ages are $\tau_{\text{WD}_\text{cool}} = [0.02, 0.14, 0.69, 2.78, 9.8]$Gyr.}
    \label{fig:WD_cooling_age}
\end{figure}

\[
{\rm log}\left[\frac{
\tau_{{\text{WD}}_{\text{cool}}}
}{\rm yr}\right] 
= 
C_1 + C_2 \ {\rm log}\left[ \frac{T_{\rm eff}}{\rm K} \right]
+
C_3  \left( {\rm log}\left[ \frac{T_{\rm eff}}{\rm K} \right]\right)^2
\]

\[
\ \ +C_4 \ {\rm exp} \left[  
\frac{-\left(\frac{T_{\rm eff}}{K}\right)^3 - \left(10^4 - \frac{T_{\rm eff}}{K} \right)^3}
{5500^3}
\right]
\cos{\left[ \frac{2\pi}{4700} \frac{T_{\rm eff}}{K} \right]}
\]

\[
\ \ +C_5 \ {\rm exp} \left[  
\frac{-\left(\frac{T_{\rm eff}}{K}\right)^3 - \left(61000 - \frac{T_{\rm eff}}{K} \right)^3}
{20000^3}
\right]
\sin{\left[ \frac{2\pi}{38000} \frac{T_{\rm eff}}{K} \right]}
\]

\begin{equation}
\end{equation}

\noindent{}such that

\begin{equation}
C_1 = 8.62 
     - 6.49 \left( \frac{M_\text{WD}}{M_{\odot}} \right)
     +13.58 \left( \frac{M_\text{WD}}{M_{\odot}} \right)^2
     - 10 \left( \frac{M_\text{WD}}{M_{\odot}} \right)^3
     ,
\end{equation}

\begin{equation}
C_2 = 3.09 
     + 2.766 \left( \frac{M_\text{WD}}{M_{\odot}} \right)
      - 5 \left( \frac{M_\text{WD}}{M_{\odot}} \right)^2
     + 3.33 \left( \frac{M_\text{WD}}{M_{\odot}} \right)^3
     ,
\end{equation}

\begin{equation}
C_3 = -0.839 
     - 0.014 \left( \frac{M_\text{WD}}{M_{\odot}} \right)
     +0.039 \left( \frac{M_\text{WD}}{M_{\odot}} \right)^2
     ,
\end{equation}

\begin{equation}
C_4 = 0.7 - 2\left|\left( \frac{M_\text{WD}}{M_{\odot}} \right) - 0.6 \right|
,
\end{equation}

\[
C_5= 250 \mathcal{H}\left[ \left( \frac{M_\text{WD}}{M_{\odot}} \right) - 0.6 \right]
+
\mathcal{H}\left[ 0.6 - \left( \frac{M_\text{WD}}{M_{\odot}} \right) \right]
\]

\begin{equation}
\ \ \ \ \times
            \left[-1910 + 6000 \left( \frac{M_\text{WD}}{M_{\odot}} \right) - 4000 \left( \frac{M_\text{WD}}{M_{\odot}} \right)^2   \right]
            ,
\end{equation}

\noindent{}where $\mathcal{H}$ is the Heaviside step function and, as is standard, log refers to log$_{10}$. This long-form relation was derived for the ranges $0.4M_{\odot} \le M_{\text{WD}} \le 0.8M_{\odot}$ and $4000 \leq T_\text{eff} \leq 30,000$K. Within these ranges, the maximum per cent error with the \cite{Fontaine2001} models is 157 per cent.

For specifically the $M_{\text{WD}} = 0.6M_{\odot}$ case, we obtain our compact relation by setting $C_4=C_5=0$ such that

\begin{equation}
    \text{log} \left[ \frac{\tau_{{\text{WD}}_{\text{cool}}}}{yr} \right] \sim 7.45 + 3.67 \:  \text{log} \left[ \frac{T_\text{eff}}{K} \right] - 0.83 \left( \text{log} \left[ \frac{T_\text{eff}}{K} \right] \right)^2.     \label{eq:cooling_age_approx}
\end{equation}

This compact relationship differs from the \cite{Fontaine2001} models with a maximum per cent error of just 25 per cent. We plot the relationship in Fig.~\ref{fig:WD_cooling_age}, where five temperatures equidistant in log-space are shown as red circles.
The points marked relate to the following effective temperatures and cooling ages $T_\text{eff} = [30,000, 18,200, 11,000, 6700, 4000]$ K and $\tau_{\text{WD}_\text{cool}} = [0.02, 0.15, 0.69, 2.78, 9.8]$ Gyr. 

Because $\tau_{{\text{WD}}_{\text{cool}}} \sim 0$ yr relates to an effective temperature of $\sim 10^5$ K, the destructive forces of the stellar radiation will have its maximum reach at this time. Observationally,  polluted white dwarfs have been observed with temperatures up to $27,000$ K \citep{Koester2014}.
Simulations have shown that disrupted material shouldn't make its way to the surface of the white dwarf until cooling ages of at least 10s of Myr \citep{Mustill2018}. The young white dwarf WD~J0914+1914, with $\tau_{{\text{WD}}_{\text{cool}}} \sim 13.3$ Myr, does not yet contain rocky pollutants despite it hosting a planetary system \citep{Gansicke2019}.
Thus it is important to consider a wide range of white dwarf temperatures and cooling ages when looking  at pollution pathways.

\subsection{Asteroid properties} \label{subsec:asteroid_props}

\subsubsection{Shape models} \label{subsubsec:shape_models}
The predecessor to this paper, BVG17, used a quasi-spherical shape model characterised by a single mean dimension $a$ and an approximate volume of $a^3$. 
However, given that Solar System asteroids have been observed with a large variety of shapes \citep{Warner2009, Durech2018}, it is likely that using such a simplified shape model affects the reliability of the results.

\begin{figure}
    \begin{subfigure}[b]{.5\linewidth}
        \centering
        \includegraphics[scale = 0.19]{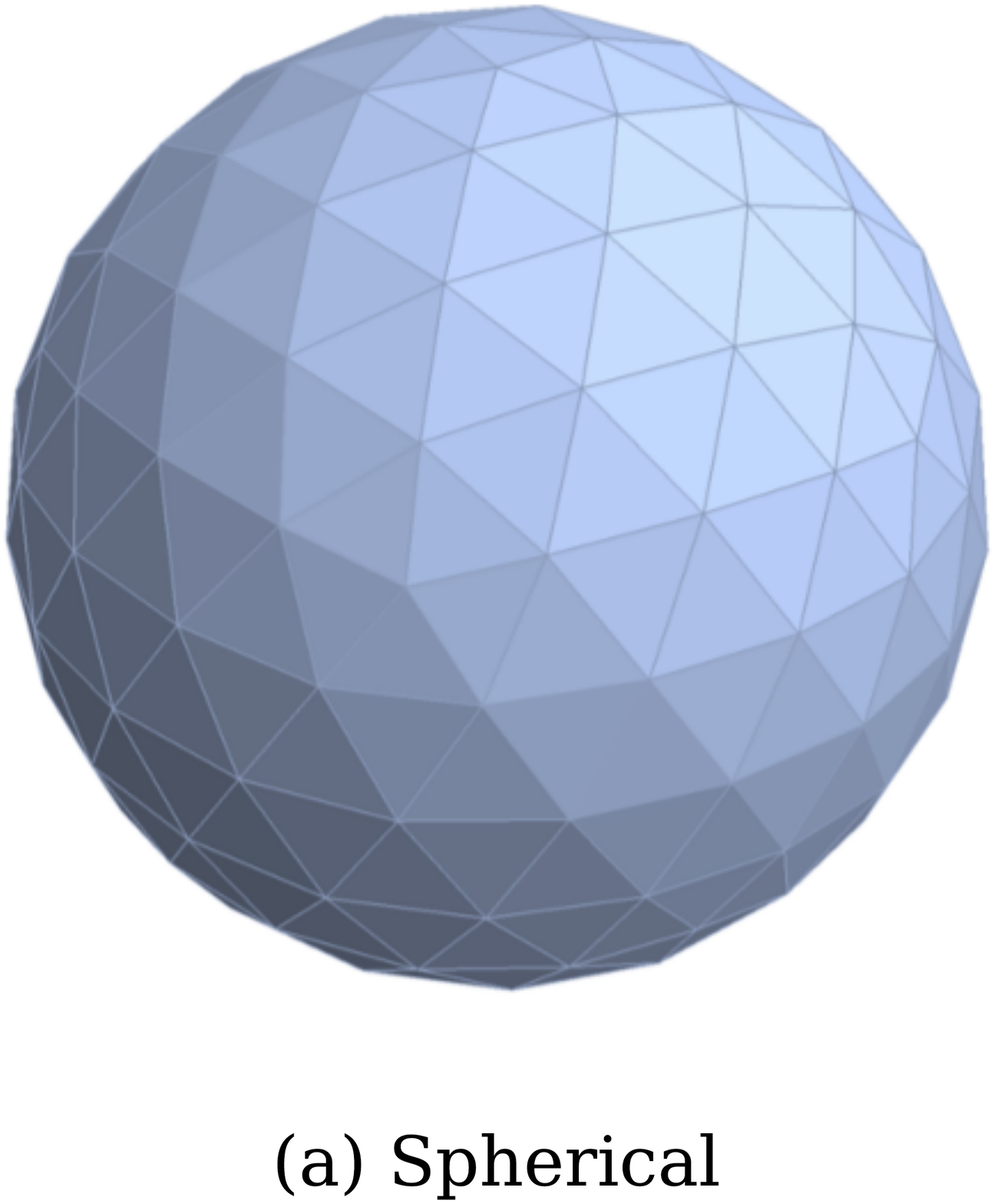}
    \end{subfigure}%
    \begin{subfigure}[b]{.5\linewidth}
        \centering
        \includegraphics[scale = 0.19]{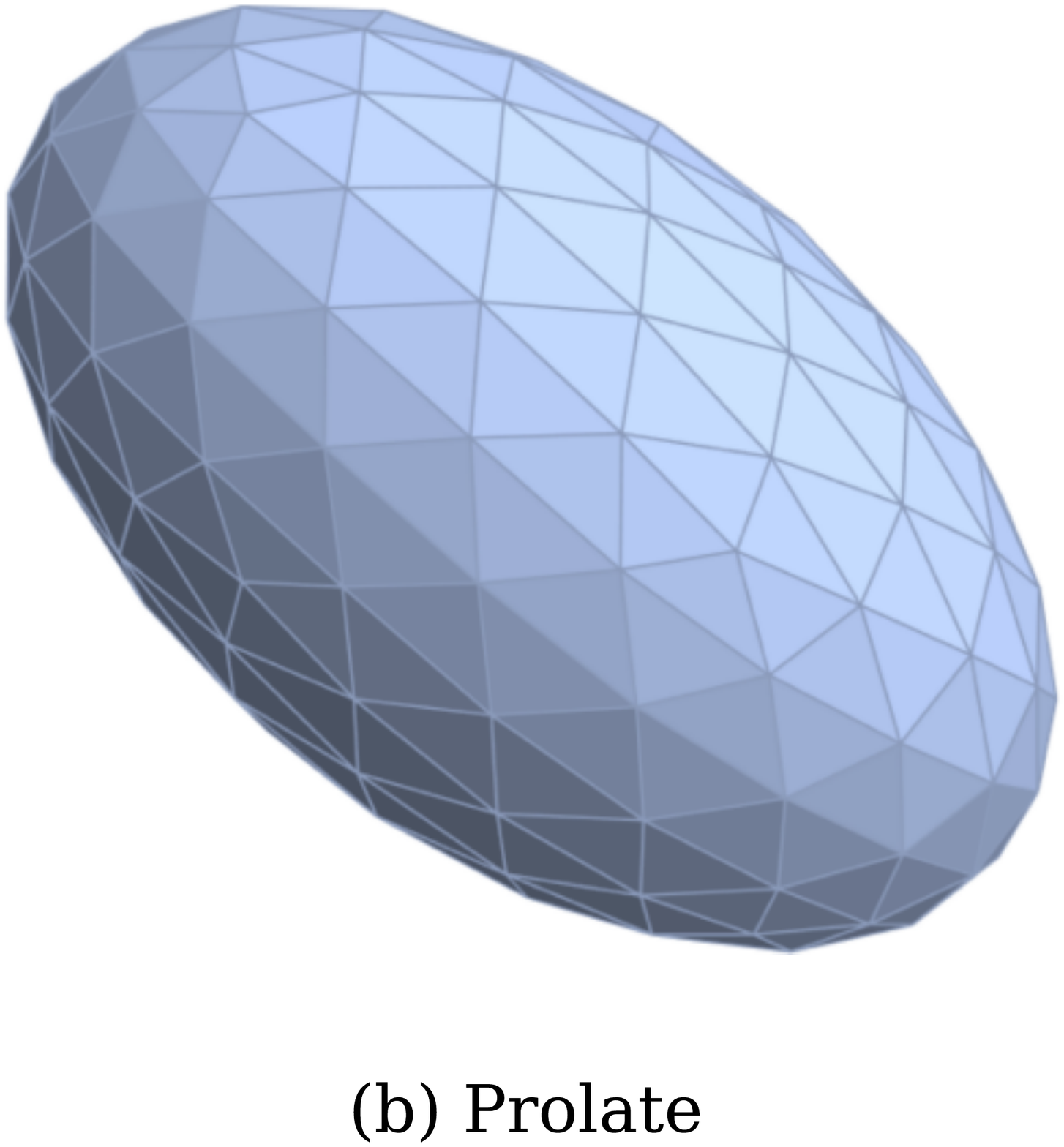}
    \end{subfigure}
    \begin{subfigure}[b]{.5\linewidth}
        \centering
        \includegraphics[scale = 0.19]{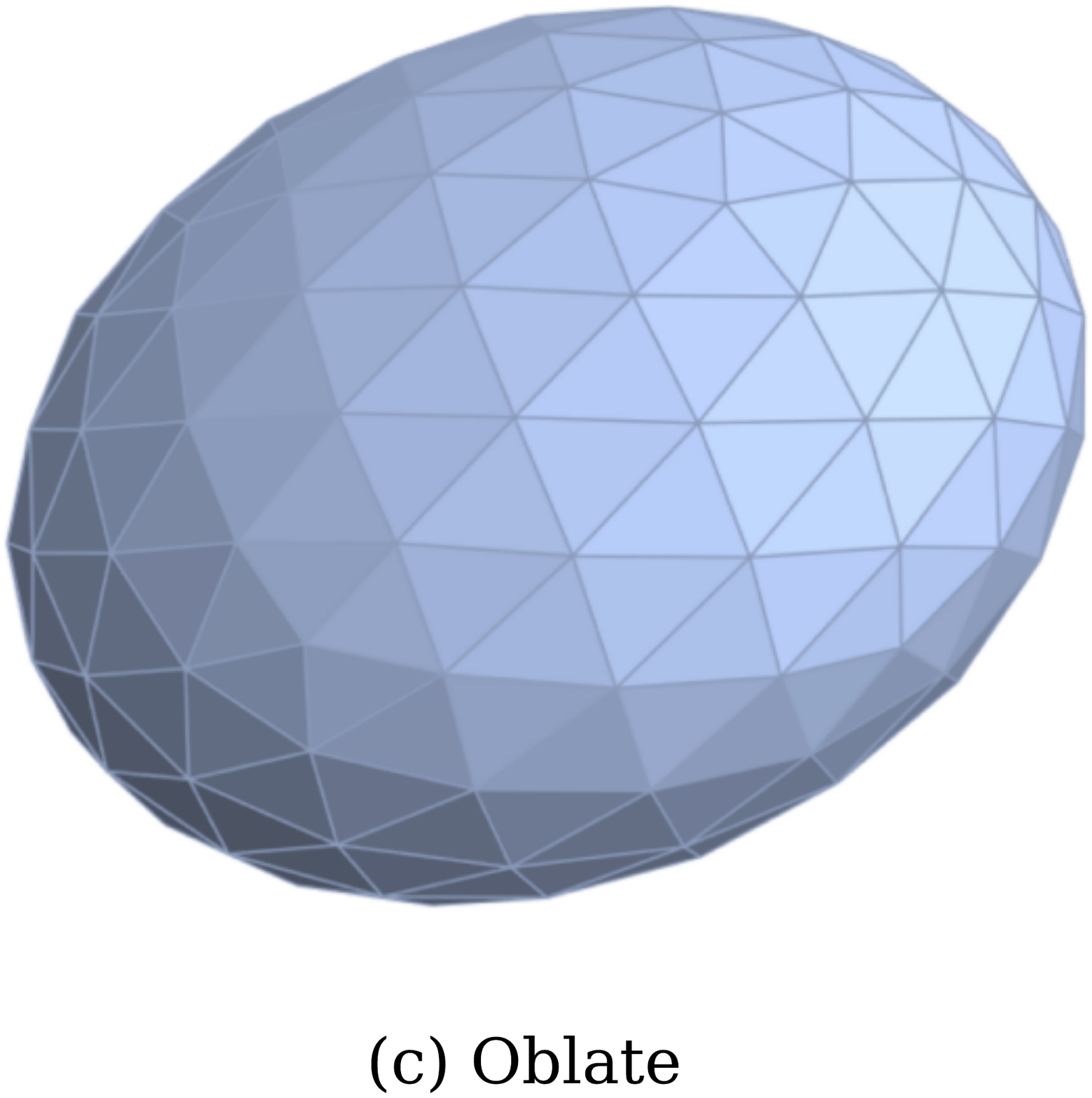}
    \end{subfigure}
    \begin{subfigure}[b]{.5\linewidth}
        \centering
        \includegraphics[scale = 0.19]{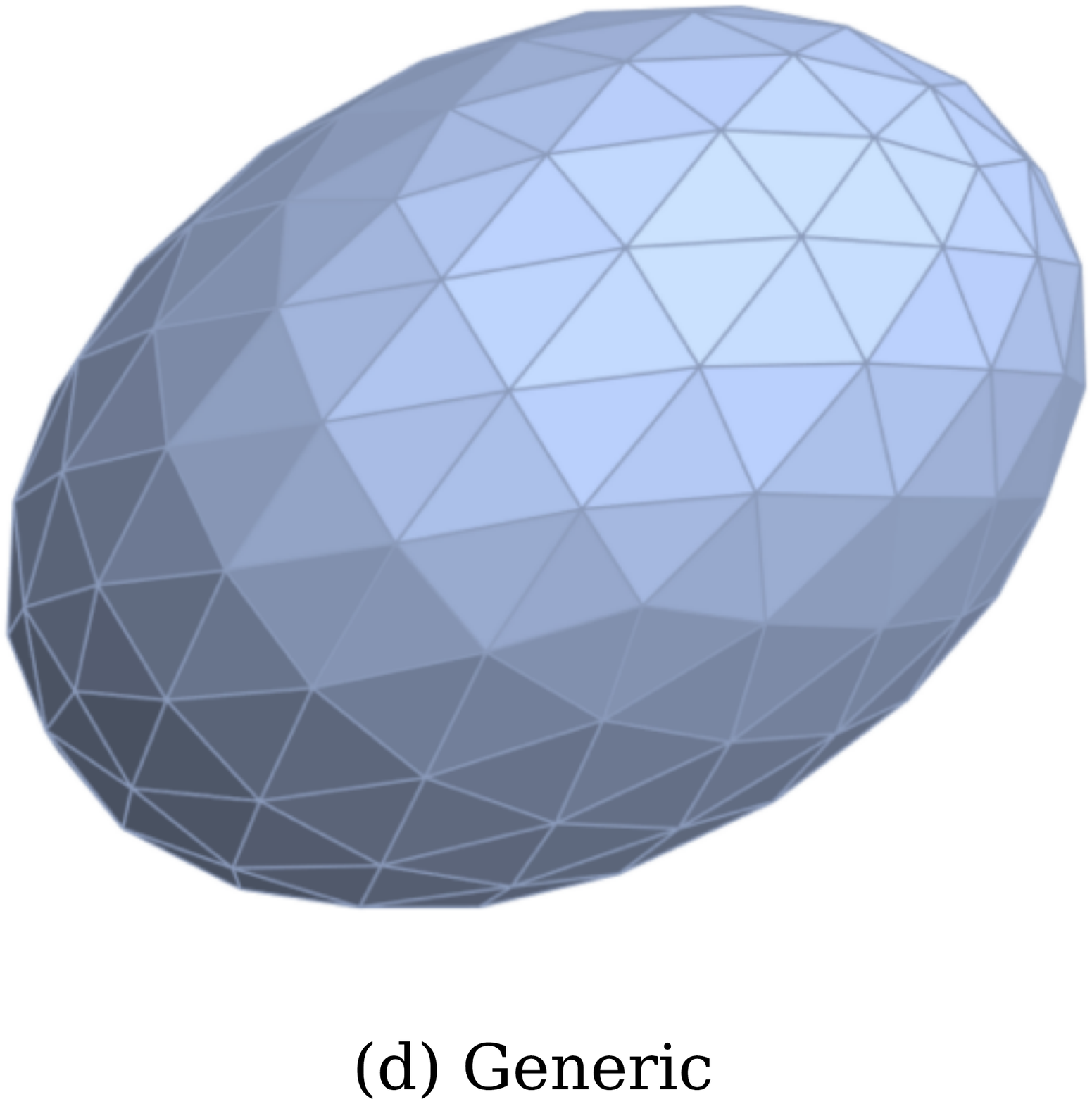}
    \end{subfigure}
    \begin{subfigure}[b]{.5\linewidth}
        \centering
        \includegraphics[scale = 0.19]{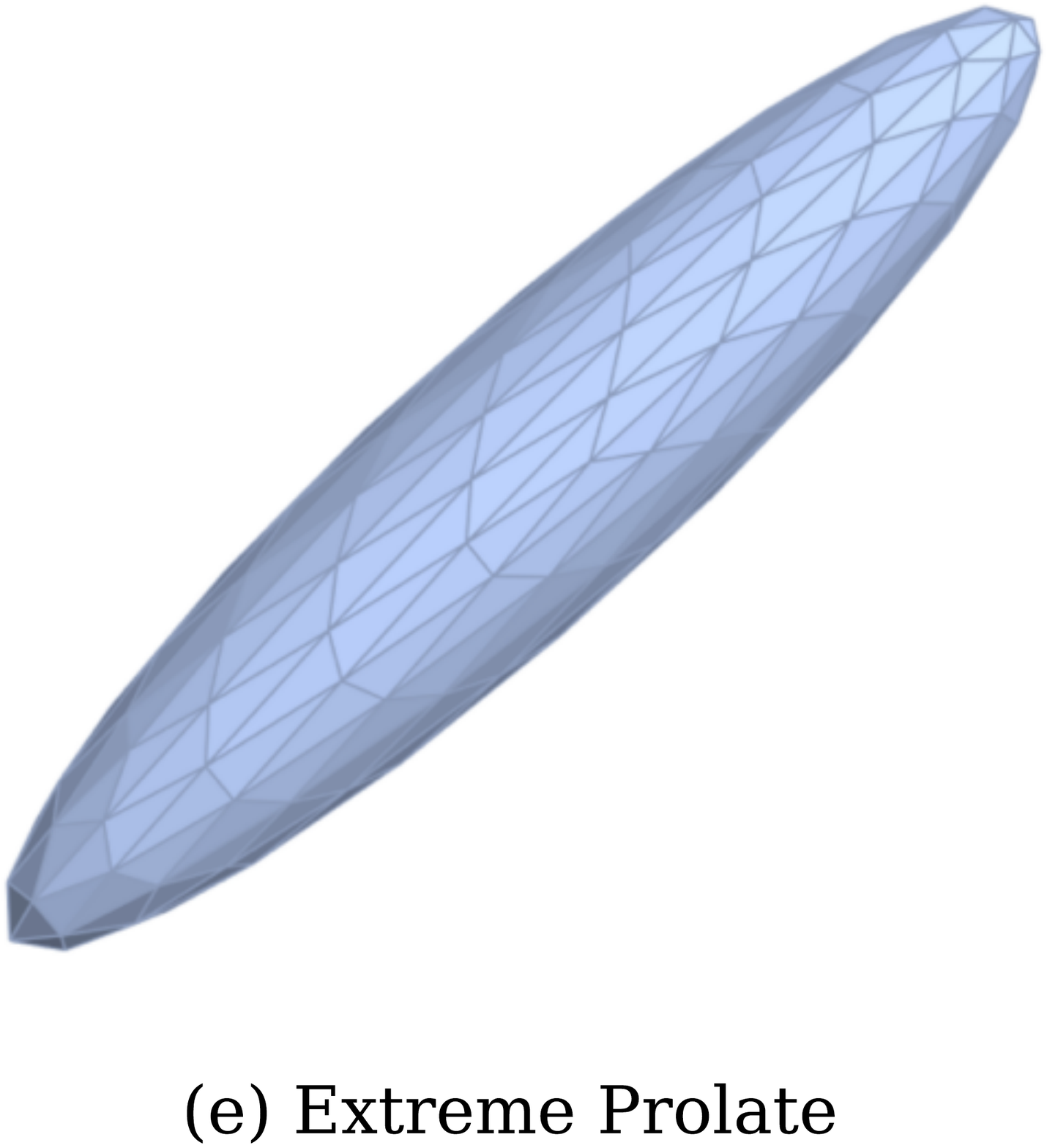}
    \end{subfigure}
    \begin{subfigure}[b]{.5\linewidth}
        \centering
        \includegraphics[scale = 0.19]{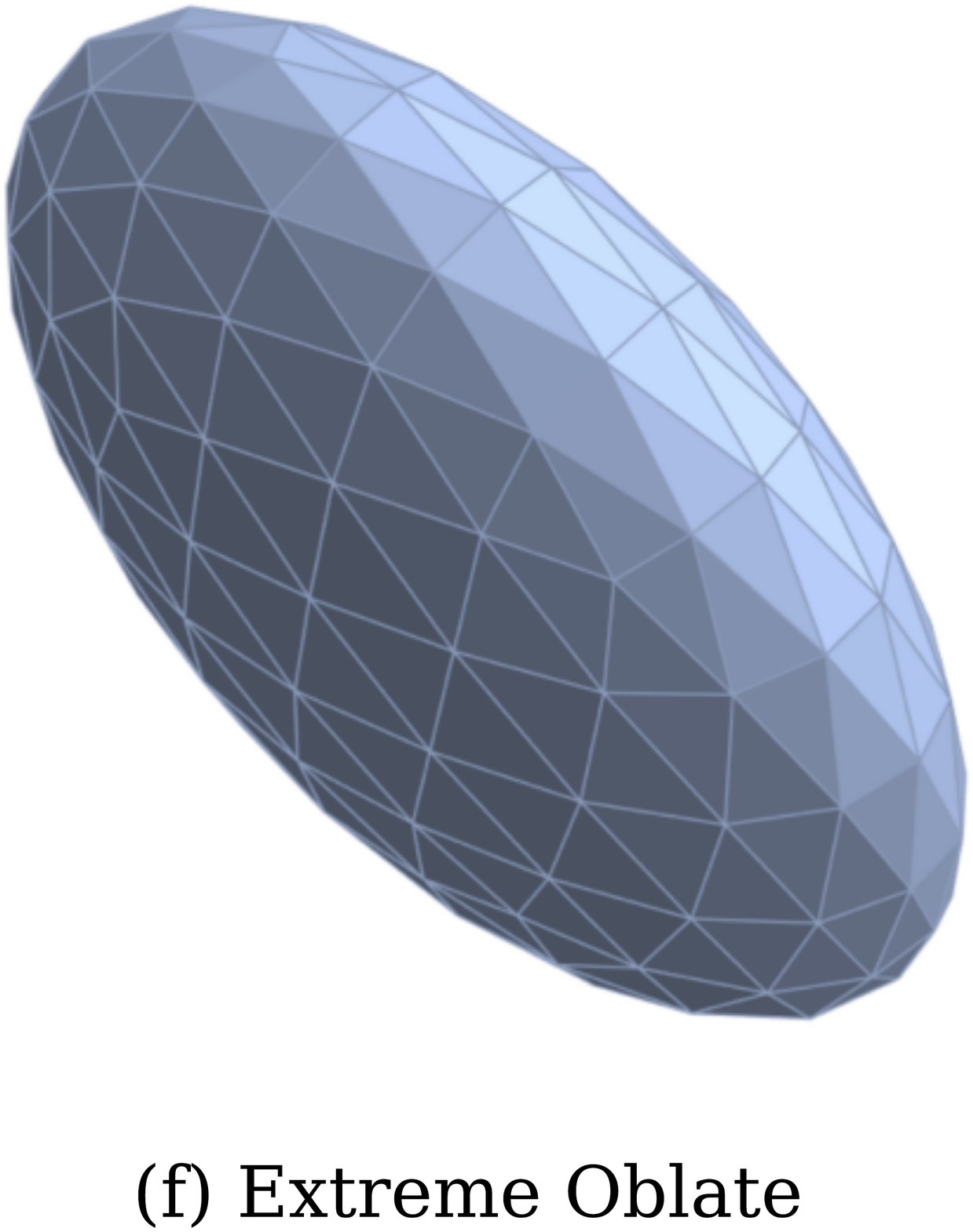}
    \end{subfigure}
    \begin{subfigure}[b]{\linewidth}
        \centering
        \includegraphics[scale = 0.19]{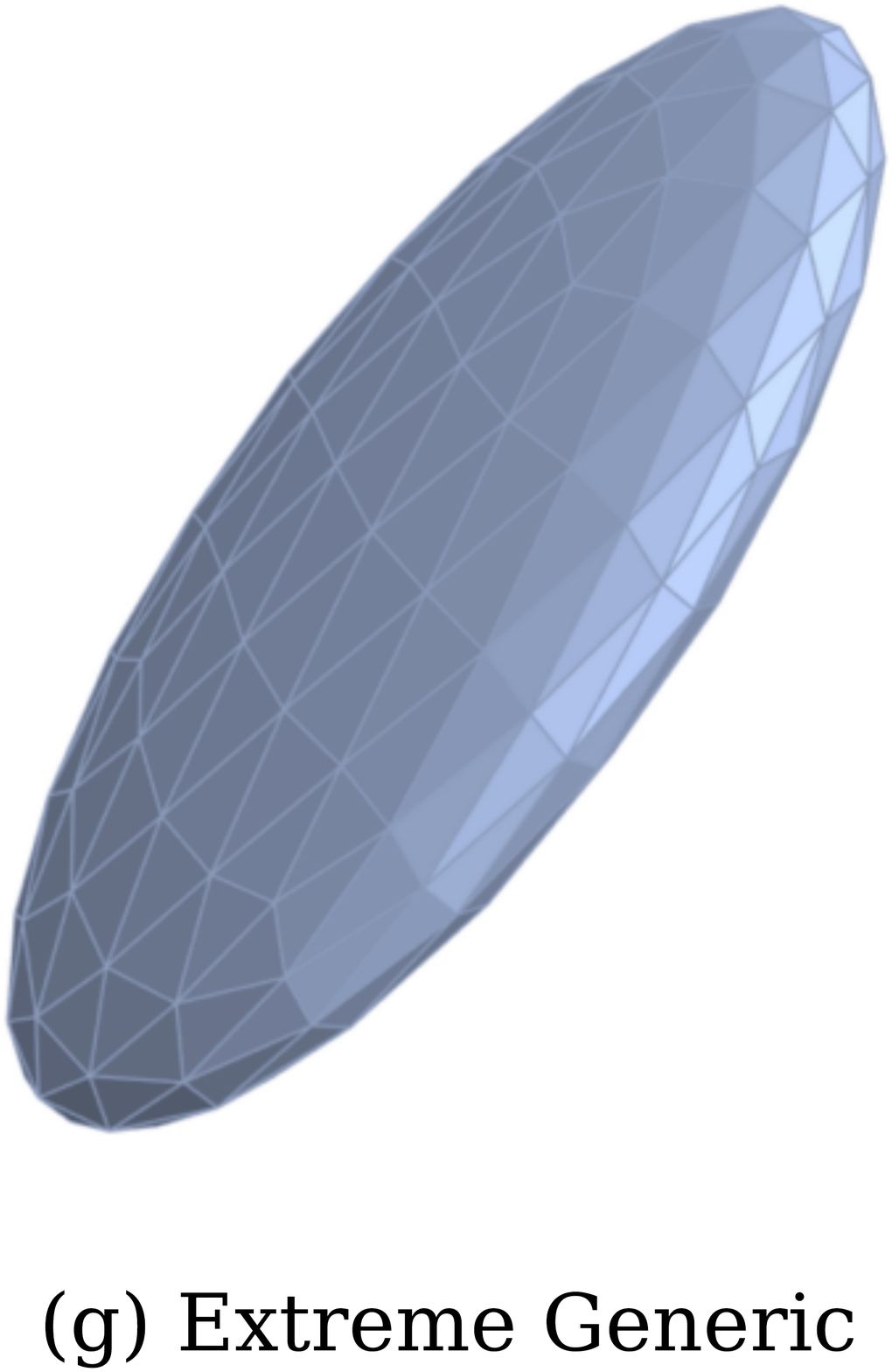}
    \end{subfigure}
    \caption{Graphical examples of the triaxial shape models detailed in Table~\ref{tab:ellip_shape_models}.}
    \label{fig:shape_models}
\end{figure}

Here, we introduce an ellipsoidal shape model characterised by the lengths of the semi-major, -intermediate and -minor axes denoted as $a$, $b$ and $c$ respectively. 
We further introduce the following aspect ratios 
\begin{equation}
    \mathfrak{b}  \equiv \frac{b}{a},
    \label{eq:b_ratio}
\end{equation}
\begin{equation}
    \mathfrak{c} \equiv \frac{c}{a},
    \label{eq:c_ratio}
\end{equation}
which allow us to focus on the shape of the ellipsoid.
Typically, asteroids are defined as oblate when $\mathfrak{b} = 1 \ne \mathfrak{c}$, and prolate when $\mathfrak{b} = \mathfrak{c} \ne a$ \citep{Holsapple2001}.
Observations of Solar System asteroids allow us to place constraints on the possible values of the aspect ratios.
Oblate asteroids have only been observed with $\mathfrak{c} > 0.4$, but prolate asteroids have been observed with $\mathfrak{c}$ values as small as 0.2 \citep{Zhang2020}. 
We consider a range of shape models which are detailed in Table~\ref{tab:ellip_shape_models} and graphically represented in Fig.~\ref{fig:shape_models}.
\begin{table}
	\centering
	\caption{Aspect ratios, $\mathfrak{b}$, $\mathfrak{c}$ for the individual shape models used in this paper.}
	\label{tab:ellip_shape_models}
	\begin{tabular}{lccr}
		\hline
		 & $\mathfrak{b}$ & $\mathfrak{c}$ & Figure\\
		\hline
		Spherical & 1.0 & 1.0 & \ref{fig:shape_models}(a)\\
		Prolate & 0.6 & 0.6 & \ref{fig:shape_models}(b)\\
		Oblate & 0.6 & 1 & \ref{fig:shape_models}(c)\\
		Generic & 0.6 & 0.8 & \ref{fig:shape_models}(d)\\
		Extreme Prolate & 0.2 & 0.2 & \ref{fig:shape_models}(e)\\
		Extreme Oblate & 0.4 & 1 & \ref{fig:shape_models}(f)\\
		Extreme Generic & 0.2 & 0.8 & \ref{fig:shape_models}(g)\\
		\hline
	\end{tabular}
\end{table}

\subsubsection{Density, porosity and mass}  \label{subsubsec:ast_density} 
For simplicity we assume that the asteroids have a uniform density throughout their bodies.

Estimates of the density of Solar System bodies derived via direct and indirect observational methods range from $\rho < 1$~\gcm for icy cometary bodies to $\rho \sim 11$~\gcm for rocky-iron bodies, although these extreme values are likely unphysical and have large associated uncertainties \citep{Carry2012}. 
To take into account this wide range of possible densities, we loosely define three material types, compared to the two used in BVG17:
\begin{enumerate}
    \item \textit{Snowy bodies:} bodies comprised of a mixture of ices and dust with a given density of $\rho_\text{Snow} = 0.5$~\gcm.
    \item \textit{Rocky bodies:} more solidified bodies made of solid ices or dust, with a given fiducial density of $\rho_\text{Rock} = 3$~\gcm.
    \item \textit{Iron bodies:} solid bodies rich in iron, with a given density of $\rho_\text{Iron} = 8$~\gcm.
\end{enumerate}

Porosity is related to the number of voids in a body that are larger than a typical micrometer crack size for meteorites \citep{Carry2012}.
Very massive, monolithic asteroids seem to have no porosity. 
Rubble pile asteroids, which are common in the Solar System, have up to 20 per cent porosity, with icier bodies exceeding this percentage. 
Constraining porosity values from observations is very difficult, and estimates for Solar System asteroids can vary by up to 30-40 per cent depending on the model used. 
Further, objects exist which go against the broad trends mentioned above and the specific evolutionary history of a body can affect its porosity.
Thus, the possible range of porosity values and hence density values is very large and in order to not overcomplicate our study, we elected to not consider porosity in the model presented here.

Assuming an ellipsoidal shape and a uniform density allows us to write the mass of an asteroid in terms of its aspect ratios as 
\begin{equation}
    \label{eq:asteroid_mass}
    M = \frac{4}{3} \pi a^3 \mathfrak{b} \mathfrak{c} \rho.
\end{equation}

\subsubsection{Latent heat} \label{subsubsec:ast_latentheat}
As a body approaches the strong radiative influence of a star, sublimation becomes an important process. 
The mass loss per unit area of an asteroidal body for a particular solar flux depends strongly on the latent heat of sublimation, $\mathcal{L}$.
Although constituent materials of an asteroid will have different values of latent heat, \cite{Brown2011} define the relevant value as the weighted mean of all mass components.
For the snowy bodies considered here this gives a value of $\mathcal{L}_\text{Snow} = 2.6 \times 10^{10}$~\ergg.
The values for solid bodies are higher: $\mathcal{L}_\text{Rock} = 8 \times 10^{10}$~\ergg and $\mathcal{L}_\text{Iron} = 10^{11}$~\ergg \citep{Chyba1993}.
 
\subsubsection{Internal strength} \label{subsubsec:ast_strength}
Asteroids approaching a white dwarf are also subjected to forces additional to sublimative mass loss. 
The effect of these forces depend on the body's ability to resist them, or how materially strong the body is. 
There are multiple strengths which help to hold a body together, including: tensile, shear and compressive strengths. 
Following the argument presented in BVG17, the variation between different strengths for a single material is less than the variation of specific strengths between materials and so only the tensile strength $S$ is considered in the following. 

Studies of the break up of Shoemaker-Levy 9 and the surface morphology of 67P/Churyumov-Gerasimenko suggest a tensile strength as low as $10^3$~\dynecm  for cometary material \citep{Greenberg1995, Groussin2015}. 
Laboratory experiments on Earth have been used to identify tensile strengths for various meteorite types which can be used in this study. 
Measurements of tensile strength for carbonaceous chondrite meteorites give a value of $\sim 4 \times 10^{7}$~\dynecm \citep{Pohl2020}.
C-type asteroids are the most common rocky asteroid type in the Solar System and provide the reservoir for C chondrites and so their measured tensile strengths are used in this study as a proxy for our model of a more general rocky asteroid. 

Here we adopt the following strength values, $S_\text{Snow} = 10^4$~\dynecm, $S_\text{Rock} = 10^{7}$~\dynecm and $S_\text{Iron} = 10^{9}$~\dynecm.

A summary of all material properties considered here can be seen in Table~\ref{tab:props}.

\begin{table}
	\centering
	\caption{Summary of material properties for the three types of asteroids considered here.}
	\label{tab:props}
	\begin{tabular}{cccc} 
		\hline
		 Type & Density $\rho$ & Latent heat $\mathcal{L}$ & Tensile strength $S$\\
		 & \gcm & \ergg & \dynecm \\
		\hline
		Snow & 0.5 & $2.6 \times 10^{10}$ & $10^4$ \\
		Rock & 3 & $8 \times 10^{10}$ & $10^7$ \\
		Iron & 8 & $10^{11}$ & $10^9$ \\
		\hline
	\end{tabular}
\end{table}

\subsubsection{Orbital properties} \label{subsubsec:ast_orbital}

Here we consider the situation where an individual asteroid is subject to an external perturbation, such as from a giant planet in the outer system, and approaches the white dwarf on a highly eccentric orbit with pericentre inside the photosphere of the white dwarf. 
This case is particularly interesting in light of the observations of the highly eccentric ($e \sim 0.97$) disrupting asteroid around the white dwarf ZTF~J0139+5145 \citep{Vanderbosch2020}.

As discussed in BVG17, during most of the asteroid's approach to the white dwarf, the orbit behaves as a linear parabolic orbit. 
Under this assumption, the velocity is only comprised of the radial speed, which can be written in the following form
\begin{equation}
    \label{eq:orbital_velocity}
    v(r) = \left( \frac{2 G M_\text{WD}}{r} \right)^{1/2} = v_* \left( \frac{R_\text{WD}}{r} \right)^{1/2} = v_* x^{-1/2},
\end{equation}
where $v_*$ is the stellar escape speed and $r$ is the distance between the centre of the white dwarf and the centre of the asteroid.
Here we introduce $x$ as the relative astrocentric distance as 
\begin{equation}
    \label{eq:astero_dist}
    x = \frac{r}{R_\text{WD}}, 
\end{equation}
to maintain notational consistency with BVG17.

\subsection{Asteroid belts} \label{subsec:belt_props}
Asteroid belts which survive giant branch evolution may be enriched by more distant icy objects via the giant branch Yarkovksy effects \citep{Veras2019} and persist as reservoirs for white dwarf pollution. 
A large portion of this matter reservoir could then be nonuniformly perturbed towards the white dwarf over a period of time ranging from orbital to Gyr timescales by post-mass-loss planetary system instability \citep{Mustill2018}.

Here we construct a simplified Main belt analogue of 100 bodies. 
Each body's shape and material is randomly chosen from those presented in Tables~\ref{tab:ellip_shape_models} and \ref{tab:props}.
Although there are almost certainly some preferences for shape and material within the Solar System's Main belt, we make the assumption here that there will be an approximate equal split between these properties. 
Much of the accreted material measured in white dwarfs appear to be from rocky, terrestrial asteroids.
However, a small number of white dwarfs have been observed with volatile and excess hydrogen pollution \citep{Farihi2013, Raddi2015, GentileFusillo2017, Xu2017, Hoskin2020}, which points to the progenitor material being water-enriched. 
Observations of the Main belt dwarf planet Ceres indicate a high polar concentration of water ice and a subsurface water ocean \citep{Prettyman2017}. 
Such water-rich bodies can persist throughout the giant branch phases of stellar evolution. \cite{Malamud2017a} find that larger bodies ($a > 25$ km) around $0.6M_\odot$ white dwarfs can retain around 50 per cent of their water content at distances of $\sim 100$ au. 
Additionally, bodies can be dynamically exchanged between the icy outer reaches of the Solar System's Kuiper belt and Oort Cloud and the rockier, terrestrial Main belt \citep{Weissman1997, Shannon2015}.
Therefore, using a mixture of rocky and snowy bodies in our analogue Main belt is realistic. 

We use an observational size distribution with a power law slope of $n \sim 0.9$ for asteroids with largest semi-axis $a > 1$~km from the High Cadence Transient Survey (HiTS) \citep{Pena2020} and slope $n \sim 0.26$ for $a < 1$~km from the Subaru Main Belt Asteroid Survey (SMBAS) \citep{Yoshida2007}.
We force 90 per cent of the 100 asteroids to be smaller than 1 km and the remaining 10 per cent to be larger. 
All sizes are then randomly chosen from the respective power law distribution.

Helium dominated DB white dwarfs have much longer diffusion timescales ($\sim$~Myr) than hydrogen dominated DAs (days to weeks), thus the level of accreted material visible in their convective zones can act as a tracer to the total amount of accreted material across the last $\sim$Myr of the planetary systems evolution and a proxy for the average across all white dwarf types.
Most polluted DB white dwarfs are estimated to have accreted $10^{21}-10^{23}$~g of planetary material \citep{Farihi2010, Xu2012, Girven2012, Veras2016a} in the last Myr, comparable with the mass of the Solar System's Main belt. 

A number of studies which utilised numerical simulations to identify how often minor bodies in white dwarf planetary systems are tidally disrupted and should contribute to the accreted material, find that in order for the fraction of their belts which accrete onto the star to match the observed totals, the overall disc mass must be several times greater than the mass of our own Main Belt \citep{Debes2012, Frewen2014}.
However, the polluted white dwarfs we currently observe largely have main sequence progenitors which were more massive than our Sun \citep{Tremblay2016, Cummings2018, ElBadry2018, McCleery2020, BarrientosChaname2021} and so persisted on the main sequence for less time, both of which could correspond with more massive belts.
While the specific dynamical evolution of the Solar System could have depleted the reservoir of asteroids in the Main belt \citep{Walsh2011}, and thus may not represent the mass of a `typical' asteroid belt.
Although the mass of the belts is undoubtedly important, and each individual asteroid has a mass, we do not ensure that the total mass in the Main belt analogue is equivalent across our simulations. 

\section{Asteroids approaching the white dwarf} \label{sec:approach}

Having laid out the physical and orbital properties of white dwarfs and asteroids, we now proceed to consider possible destruction regimes for an ellipsoidal body approaching a white dwarf on a linear trajectory. We assume that the scatterer is the only or innermost major planet in the system, such that nothing generates deviations in this trajectory.

We consider three possible regimes: \textit{i) sublimation} where the bodies total mass is lost due to the incident energy flux; \textit{ii) fragmentation} where tidal forces from the star overcome the body's own internal tensile strength and gravitational forces; and \textit{iii) impact} where the body enters the white dwarf photosphere and undergoes frictional ablative mass loss and/or ram-pressure pancaking and deceleration effects.

Much of the following analysis in this paper proceeds in much the same way as presented in BVG17, however with the introduction of considering the three principle directions of the three dimensional ellipsoidal model separately.

In the following we consider the cartesian basis \{${\hat{i}, \hat{j}, \hat{k}}$\} to be aligned with the ellipsoid's semi-axes such that $\hat{i}$ and $\hat{j}$ are aligned with the largest and intermediate semi-axes. 
This allows us to define the cross-sectional areas of the asteroids in the three principal directions in terms of the largest semi-axis $a$ and the aspect ratios $\mathfrak{b}$ and $\mathfrak{c}$ as
\begin{equation}
    \label{eq:areas}
    \text{cross-sectional area} = \left\{ \pi a^2 \mathfrak{b} \mathfrak{c}\hat{i}, \: \pi a^2 \mathfrak{c} \hat{j}, \: \pi a^2 \mathfrak{b} \hat{k} \right\}.
\end{equation}

\subsection{Sublimation} \label{subsec:sublimation}
Outside the Roche limit of the white dwarf, sublimation should dominate the disruption of infalling planetesimals and is governed by the incident flux of starlight on a body.

Assuming the incoming asteroid has near-zero albedo the incident power of the starlight $\mathbfit{P}_*$ is
\begin{equation}
    \label{eq:incident_power}
    \mathbfit{P}_* = \frac{a^2 \pi \sigma T_*^4}{x^2} \left[\mathfrak{b} \mathfrak{c} \hat{i} + \mathfrak{c} \hat{j} + \mathfrak{b} \hat{k} \right].
\end{equation}

By considering the mass loss from the body due to this incident starlight, in the same way as in BVG17, we can find an expression for how the largest semi-axis varies as a function of astrocentric distance due to sublimative effects
\begin{equation}
    \label{eq:a_sub}
    a_{\text{sub}}(x) = a_0 - \frac{\mathbfit{A}}{x^{1/2}}.
\end{equation}
$\mathbfit{A}$ is the \textit{sublimation parameter,}
\begin{equation}
    \label{eq:sub_param}
    \mathbfit{A} = \frac{R_\text{WD}^{3/2} T_\text{eff}^4 \sigma}{2\sqrt{2GM_\text{WD}} \mathcal{L} \rho} \left[ \hat{i} + \frac{1}{\mathfrak{b}} \hat{j} + \frac{1}{\mathfrak{c}}\hat{k} \right],
\end{equation}
and represents the minimum value of the largest semi-axis size scaled by the relative astrocentric distance which allows an asteroid to withstand sublimative forces alone until it reaches the white dwarf photosphere. $a_0$ is the initial size of the asteroid.
This value differs from that presented in BVG17 by a numerical factor in the $x$-direction and additional factors of the aspect ratios in the $y$ and $z$-directions.

This sublimation model neglects cooling effects, assumes that the sublimation occurs much more quickly than radiative energy is otherwise lost and does not take into account the intrinsic vapour pressure, interactions with an extant accretion disc or other effects which might alter the sublimation process.
See \cite{Steckloff2015} for a more complete treatment of the sublimation process and \cite{Steckloff2021} for said treatment applied around white dwarfs.
Further details about the derivation of this value can be found in Appendix~\ref{app:sub} and BVG17.

\subsection{Fragmentation} \label{subsec:fragmentation}
A minor body will fragment into smaller child bodies when the tidal forces from the white dwarf overcome the body's own internal forces.
Here we will simply consider the body's self-gravitational and tensile strength forces as interior forces. 
While this approach is similar to the process in BVG17, we again introduce three dimensional ellipsoidal forces.

The tidal force on an ellipsoidal body due to a large central body is
\begin{equation}
    \label{eq:force_tidal}
    \mathcal{\mathbfit{F}}_T = \frac{G M M_\text{WD} a}{x^3 R_\text{WD}^3} \left[ 2 \hat{i} - \mathfrak{b} \hat{j} - \mathfrak{c} \hat{k} \right].
\end{equation}
This form assumes that the body rotates about the $z$-axis -- which has the greatest inertia -- and that the $x$-axis, with the least inertia, always points towards the central body. This form also assumes that librations about this orientation can be neglected \citep{Dobrovolskis2019}. 
This assumption is further discussed in Section~\ref{sec:rotation}.

The force due to the internal tensile strength is the cross-sectional area (equation~\ref{eq:areas}) multiplied by the material's tensile strength \citep{Davidsson2001}, 
\begin{equation}
    \label{eq:force_tensile}
    \mathcal{\mathbfit{F}_S} = - \pi a^2 S \left[ \mathfrak{c} \mathfrak{b} \hat{i} + \mathfrak{c} \hat{j} + \mathfrak{b} \hat{k} \right].
\end{equation}
For a homogeneous ellipsoid, such as considered here, the self-gravitational force depends only on the shape of the asteroid through the aspect ratios and can be written as follows \citep{Holsapple2004, Holsapple2006},
\begin{equation}
    \label{eq:force_selfgrav}
    \mathcal{\mathbfit{F}_G} = - 2 \pi G M \rho a \left[ U_x \hat{i} + U_y \mathfrak{b} \hat{j} + U_z \mathfrak{c} \hat{k} \right],
\end{equation}
with
\begin{equation}
    \label{eq:Ax}
    U_x = \mathfrak{b}\mathfrak{c} \int^\infty_0 \frac{du}{\left( u + 1\right)^{3/2}\left(u + \mathfrak{b}^2\right)^{1/2} \left(u + \mathfrak{c}^2\right)^{1/2}},
\end{equation}
\begin{equation}
    \label{eq:Ay}
    U_y = \mathfrak{b}\mathfrak{c} \int^\infty_0 \frac{du}{\left( u + 1\right)^{1/2}\left(u + \mathfrak{b}^2\right)^{3/2} \left(u + \mathfrak{c}^2\right)^{1/2}},
\end{equation}
\begin{equation}
    \label{eq:Az}
    U_z = \mathfrak{b}\mathfrak{c} \int^\infty_0 \frac{du}{\left( u + 1\right)^{1/2}\left(u + \mathfrak{b}^2\right)^{1/2} \left(u + \mathfrak{c}^2\right)^{3/2}}.
\end{equation}
The strength and gravitational forces together resist the influence of the tidal force.

The net force acting on the triaxial body is then 
\begin{equation}
    \label{eq:force_net}
    \mathcal{\mathbfit{F}_{\text{tot}}} = + |\mathcal{\mathbfit{F}_T}| - |\mathcal{\mathbfit{F}_S}| - |\mathcal{\mathbfit{F}_G}|
\end{equation}
and the condition for a body to resist the tidal forces of the central star is 
\begin{equation}
    \label{eq:cond_resist_frag}
    \frac{| \mathcal{\mathbfit{F}_S} + \mathcal{\mathbfit{F}_G} | }{\mathcal{\mathbfit{F}_T}} \gtrsim 1.
\end{equation}

In this paper, we want to focus on the case where internal strength dominates over the self-gravity of the body, which is likely to be the case for many of the smaller bodies assumed here.
Although `rubble pile' asteroids dominated by self-gravity are expected to be common in the Solar System, they are likely to break into constituent particles through tidal disruption or rotational fission. 
The result would be a large collection of smaller bodies with much higher internal integrity than the parent body, dominated by internal strength. 

The size where a body's internal strength dominates over self-gravity, occurs when the ratio between the internal strength force (equation~\ref{eq:force_tensile}) and self-gravitational force (equation~\ref{eq:force_selfgrav}) is greater than 1 as follows, 
\begin{equation}
    \frac{\mathcal{\mathbfit{F}_S}}{\mathcal{\mathbfit{F}_G}} > 1.
    \label{eq:force_ratio}
\end{equation}
\begin{table*}
	\centering
	\caption{The ratio of internal strength to self gravitation forces (equation~\ref{eq:force_ratio}) for all of the shape models and compositions considered in this paper. The values where the assumption that internal strength dominates over self-gravity breaks down are highlighted in red.}
	\label{tab:force_ratio}
	\begin{tabular}{ccccccccccccccccccc} 
		\hline
		 Size & \multicolumn{3}{c}{Generic} & \multicolumn{3}{c}{Extreme Generic} & \multicolumn{3}{c}{Prolate} & \multicolumn{3}{c}{Extreme Prolate} & \multicolumn{3}{c}{Oblate} & \multicolumn{3}{c}{Extreme Oblate} \\
		 (cm) & Snow & Rock & Iron & Snow & Rock & Iron & Snow & Rock & Iron & Snow & Rock & Iron & Snow & Rock & Iron & Snow & Rock & Iron  \\
		\hline
		$10^{3}$ & $10^{5}$ & $10^{6}$ & $10^{7}$ & $10^{5}$ & $10^{6}$ & $10^{8}$ & $10^{5}$ & $10^{6}$ & $10^{7}$ & $10^{5}$ & $10^{7}$ & $10^{8}$ & $10^{5}$ & $10^{6}$ & $10^{7}$ & $10^{5}$ & $10^{6}$ & $10^{7}$ \\
		$10^{4}$ & $10^{3}$ & $10^{4}$ & $10^{5}$ & $10^{3}$ & $10^{4}$ & $10^{6}$ & $10^{3}$ & $10^{4}$ & $10^{5}$ & $10^{3}$ & $10^{5}$ & $10^{6}$ & $10^{3}$ & $10^{4}$ & $10^{5}$ & $10^{3}$ & $10^{4}$ & $10^{5}$ \\
		$10^{5}$ & $10^{1}$ & $10^{2}$ & $10^{3}$ & $10^{1}$ & $10^{2}$ & $10^{4}$ & $10^{1}$ & $10^{2}$ & $10^{3}$ & $10^{1}$ & $10^{3}$ & $10^{4}$ & $10^{1}$ & $10^{2}$ & $10^{3}$ & $10^{1}$ & $10^{2}$ & $10^{3}$ \\
		$10^{6}$ & \red{$10^{-1}$} & $10^{0}$ & $10^{1}$ & \red{$10^{-1}$} & $10^{0}$ & $10^{2}$ & \red{$10^{-1}$} & $10^{0}$ & $10^{1}$ & \red{$10^{-1}$} & $10^{1}$ & $10^{2}$ & \red{$10^{-1}$} & $10^{0}$ & $10^{1}$ & \red{$10^{-1}$} & $10^{0}$ & $10^{1}$ \\
		$10^{7}$ & \red{$10^{-3}$} & \red{$10^{-2}$} & \red{$10^{-1}$} & \red{$10^{-3}$} & \red{$10^{-2}$} & $10^{0}$ & \red{$10^{-3}$} & \red{$10^{-2}$} & \red{$10^{-1}$} & \red{$10^{-3}$} & \red{$10^{-1}$} & $10^{0}$ & \red{$10^{-3}$} & \red{$10^{-2}$} & \red{$10^{-1}$} & \red{$10^{-3}$} & \red{$10^{-2}$} & \red{$10^{-1}$} \\
		$10^{8}$ & \red{$10^{-5}$} & \red{$10^{-4}$} & \red{$10^{-3}$} & \red{$10^{-5}$} & \red{$10^{-4}$} & \red{$10^{-2}$} & \red{$10^{-5}$} & \red{$10^{-4}$} & \red{$10^{-3}$} & \red{$10^{-5}$} & \red{$10^{-3}$} & \red{$10^{-2}$} & \red{$10^{-5}$} & \red{$10^{-4}$} & \red{$10^{-3}$} & \red{$10^{-5}$} & \red{$10^{-4}$} & \red{$10^{-3}$} \\
		\hline
	\end{tabular}
\end{table*}
To quantify the boundary where the internal strength dominates, and thus where we can neglect self gravity, we calculate the ratio in equation~(\ref{eq:force_ratio}) and present an order of magnitude result for the $x$-component of the ratio in Table~\ref{tab:force_ratio}.
As is discussed further in Section \ref{subsec:outcomes}, fragmentation always occurs in the $x$-direction of the body and thus here we only consider the ratio of the $x$-components of the forces. 
The largest asteroid size considered in our study is $10^8$cm, since the assumption that the internal strength dominates over self-gravity begins to break down for the weakest comet-like bodies at $10^6$cm and the strongest iron bodies at $10^8$cm as shown in Table~\ref{tab:force_ratio}.
The extremely strong iron planetesimal inferred to be orbiting inside the Roche limit of the white dwarf SDSS~J122859.93+104032.9 (hereafter SDSS~J1228+1040) is assumed to be between $2\times 10^5 - 2 \times 10^7$cm in size \citep{Manser2019}, which puts this object firmly in the regime where internal strength dominates over self-gravitation as defined in this paper. 

A further way to examine the body size at which internal strength dominates is to look at a graph of frequency against period and diameter, such as fig. 1 of \cite{Hestroffer2019}.
In such a graph, the spin barrier (the rotational velocity at which a rubble pile asteroid will undergo rotational fission) is clearly visible.
The location of the spin barrier implies that rubble pile asteroids begin to dominate the asteroid population between $10^4-10^5$cm in size. 
Although this estimate is below the minimum size for a rubble pile body $10^6$ cm found from Table~\ref{tab:force_ratio}, fast monolithic rotators have been observed and thus our slightly larger estimate is not unreasonable. 

In this case, we can neglect self-gravity. As discussed in BVG17, unless a body is particularly large the tensile strength is much more important in resisting tidal forces than the body's self-gravitation. 
When solely considering tensile strength as a resistive force, the condition for fragmentation becomes 
\begin{equation}
    \label{eq:a_frag}
    a(x) < a_{\text{frag}}(x) = \mathbfit{B} x^{3/2}
\end{equation}
where 
\begin{equation}
    \label{eq:binding_param}
    \mathbfit{B} = \sqrt{\frac{3R_\text{WD}^3S}{4GM_\text{WD}\rho}}\left[ \frac{\hat{i}}{\sqrt{2}} + \frac{\hat{j}}{\mathfrak{b}} + \frac{\hat{k}}{\mathfrak{c}} \right]
\end{equation}
is the \textit{binding size} parameter and represents the size that must be exceeded for a body to fragment.   
Again, this only differs from the BVG17 result by a numerical factor and the aspect ratios in the semi-intermediate and -minor axes.

\subsection{Impact} \label{subsec:impact}
If the body is not completely destroyed by sublimation, and it never meets the criteria to fragment, then we assume that the body enters the photosphere of the white dwarf. 
In this subsection, we discuss qualitatively what happens to the body during impact. For a quantitative discussion see Section 6 of \cite{Brown2017}, or, for the analogous Sun-comet case, see \cite{Brown2015}.

As discussed in Section~\ref{subsubsec:ast_orbital}, the bodies considered here propagate along orbits with orbital pericentres inside of the photosphere of the white dwarf and with periastron distance $q=0$.
Once the body enters the photosphere of the white dwarf, it encounters a very different environment. 
The white dwarf's very high gravity and temperature restricts the atmosphere to have a very small density scale height. 
Due to this scale height, in the photosphere the atmospheric frictional heating very quickly overcomes the effect of any further sublimation, and ablation becomes dominant.

The ablation also has to battle with the deceleration effects felt by the body. 
The majority of the incident atmospheric bombardment flux felt by the body as it moves through the photosphere does not reach the nucleus to ablate it, but instead heats up the surrounding atmosphere through a bow shock, decelerating the body and ablating the nucleus through heat transfer. 
Under these intense forces, the body can only survive passage through a few vertical scale heights of the atmosphere.

\subsection{Outcomes} \label{subsec:outcomes}
As a body moves towards the white dwarf, its instantaneous size at any astrocentric distance $x$ is described by the sublimation size, $a_\text{sub}(x)$, as in equation~(\ref{eq:a_sub}). 
For fragmentation to occur, this instantaneous size must coincide with the fragmentation condition, $a_\text{frag}(x)$ (equation~\ref{eq:a_frag}) at the particular distance. 
Again, we only consider the $x$ components of the sublimation and binding parameters, $A_x$ and $B_x$ respectively as fragmentation only occurs in the $x$-direction.
This coincidence requires the two functions to touch at $a_\text{sub}(x) = a_\text{frag}(x)$.

\begin{figure}
	\includegraphics[width=\columnwidth]{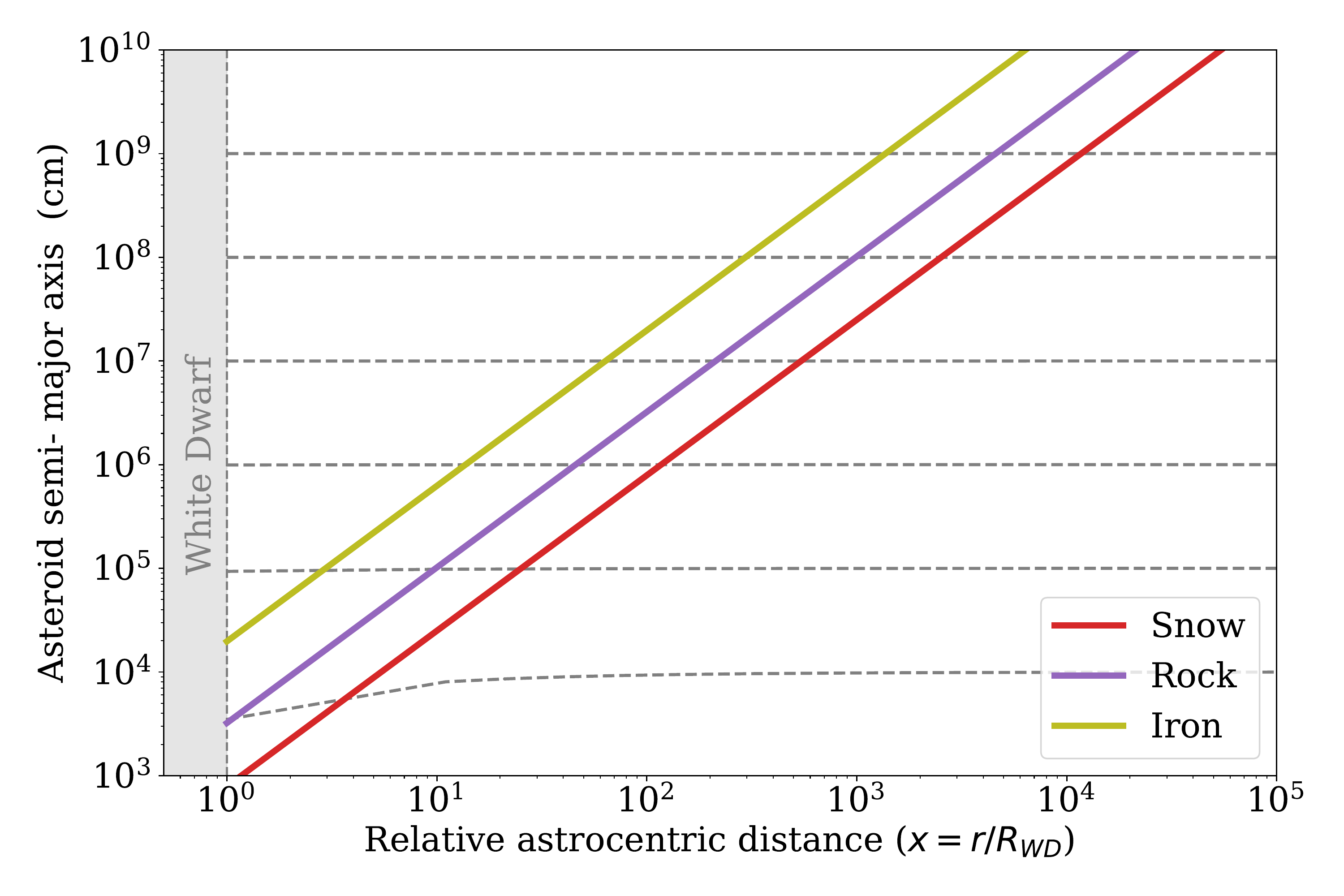}
    \caption{Asteroid fragmentation will occur when the instantaneous size of the body due to sublimation (dashed lines) coincides with the size and location condition for the body to fragment (solid coloured lines).
    The large size of these bodies resist large amounts of size changes due to sublimation and so these lines are effectively straight.
    The less tensile strength and density a body has, the further away from the white dwarf fragmentation can occur.
    As fragmentation always occurs in the $x$-direction, and the $x$-components of both the sublimation and binding size parameters are independent of shape model, this graphical representation of the fragmentation condition is valid for all shape models presented in this work.}
    \label{fig:crossing_function}
\end{figure}

The two functions are plotted in Fig.~\ref{fig:crossing_function}.
The grey dashed lines show the sublimation size for different initial asteroid sizes.
At large asteroid sizes ($a > 10^4$cm), only a small amount of the bodies total mass is lost due to sublimation, regardless of the body's material.
The instantaneous size of the body is largely independent of material and, only the weakest (snowy), smallest bodies lose mass due to sublimation.
Thus only the instantaneous size for a snowy body is shown. 
The lines for an equal sized body of rock or iron would follow the same straight trajectory as the dashed grey lines visible in Fig~\ref{fig:crossing_function}.
The coloured lines indicate the required size for fragmentation to occur for each material described in Table~\ref{tab:props}.
The points where the grey and coloured lines intersect identify the positions where fragmentation will occur.
If the condition for fragmentation is not met, the body will continue to lose mass through sublimation until there is no more mass to be lost, or the body impacts directly onto the white dwarf.

Here, we would like to visibly track the size changes of the asteroid as it approaches the white dwarf and pinpoint the moment, and mode of disruption.
Thus, we calculate the sublimation size (equation \ref{eq:a_sub}) in each semi-axis across a fine grid of astrocentric distance values until one of the destruction criterion is reached in any of the body's three principal directions. 

\begin{figure}
	\includegraphics[width=\columnwidth]{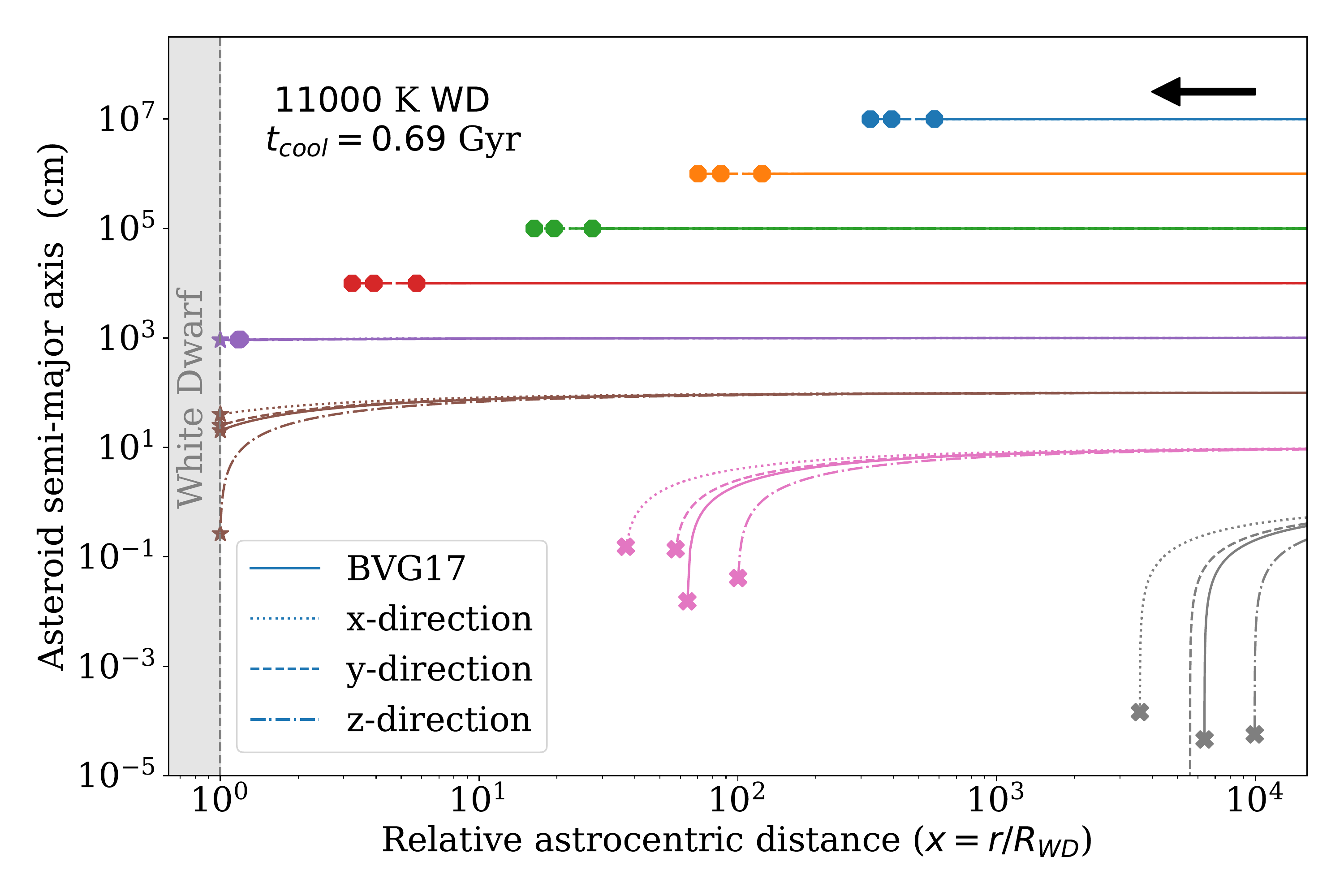}
    \caption{An example of following the body's change in size during approach and identifying destruction.
    Here, the astrocentric distance is on the $x$-axis, with the asteroid moving from right to left. The body's largest semi-axis $a$ is plotted on the $y$-axis. 
    The white dwarf temperature is chosen at $11000$K and initial sizes in the range $10^0-10^7$cm are considered.
    The body is assumed to be of snowy composition, and the generic shape model described in Table~\ref{tab:ellip_shape_models} is used.
    The line styles are as described on the figure legend.
    A circle marker indicates that the body ultimately fragments, a star indicates impact and the cross indicates that the body completely sublimates. 
    Fragmentation always occurs in the $x$-direction of the triaxial model and at the same locations as fragmentation in the BVG17 model.
    Sublimation occurs in the smallest, $z$-direction, first. Hence, the BVG17 model consistently underestimates the distance from the white dwarf where complete sublimation occurs.}
    \label{fig:comparison_plot}
\end{figure}

Fig.~\ref{fig:comparison_plot} shows an example of this process, where the asteroid moves from right to left. 
The figure shows how the size of each ellipsoidal semi-axis varies due to sublimation as the asteroid approaches the white dwarf, compared to the same size variation presented in BVG17.
In all cases where fragmentation occurs, it is always the $x$-direction (largest semi-axis) that fragments first.
As the body's least stable part is at the end of its longest axis \citep{Harris1996}, fragmentation occurring in the $x$-direction is not unexpected.
It should also be noted, that considering just the $x$-direction would be the same as considering the bodies as purely spherical with the semi-intermediate and -minor axes the same length as the semi-major axis. 
For the bodies which sublimate completely ($a < 1$cm) in Fig~\ref{fig:comparison_plot}, the smallest semi-axis ($z$-direction) consistently completely sublimates first. 
Although this failure in the $z$-direction (dot-dash line) occurs further away from the white dwarf than the case presented in BVG17 (solid line) and the equivalent purely spherical case (dotted line), this increased distance of sublimation is not large enough to drastically change the expected debris distribution from such an asteroid sublimating.

A further discussion on the effect of chosen triaxial shape model on the process of sublimation can be found in Section~\ref{subsubsec:shape_sub}.
An alternative method to identifying which type of destruction will befall a particular asteroid is discussed in Appendix~\ref{app:outcomes} and in BVG17.

\section{A Main Belt Analogue} \label{sec:main_belt}
By using the structure and properties of a Main belt analogue discussed in Section~\ref{subsec:belt_props}, here we aim to identify the fates of the bodies in such a belt if each asteroid was perturbed onto an extremely eccentric ($e \sim 1$), effectively linear, parabolic orbit.

\begin{figure*}
    \begin{subfigure}[c]{.5\linewidth}
        \centering
        \includegraphics[scale = 0.299]{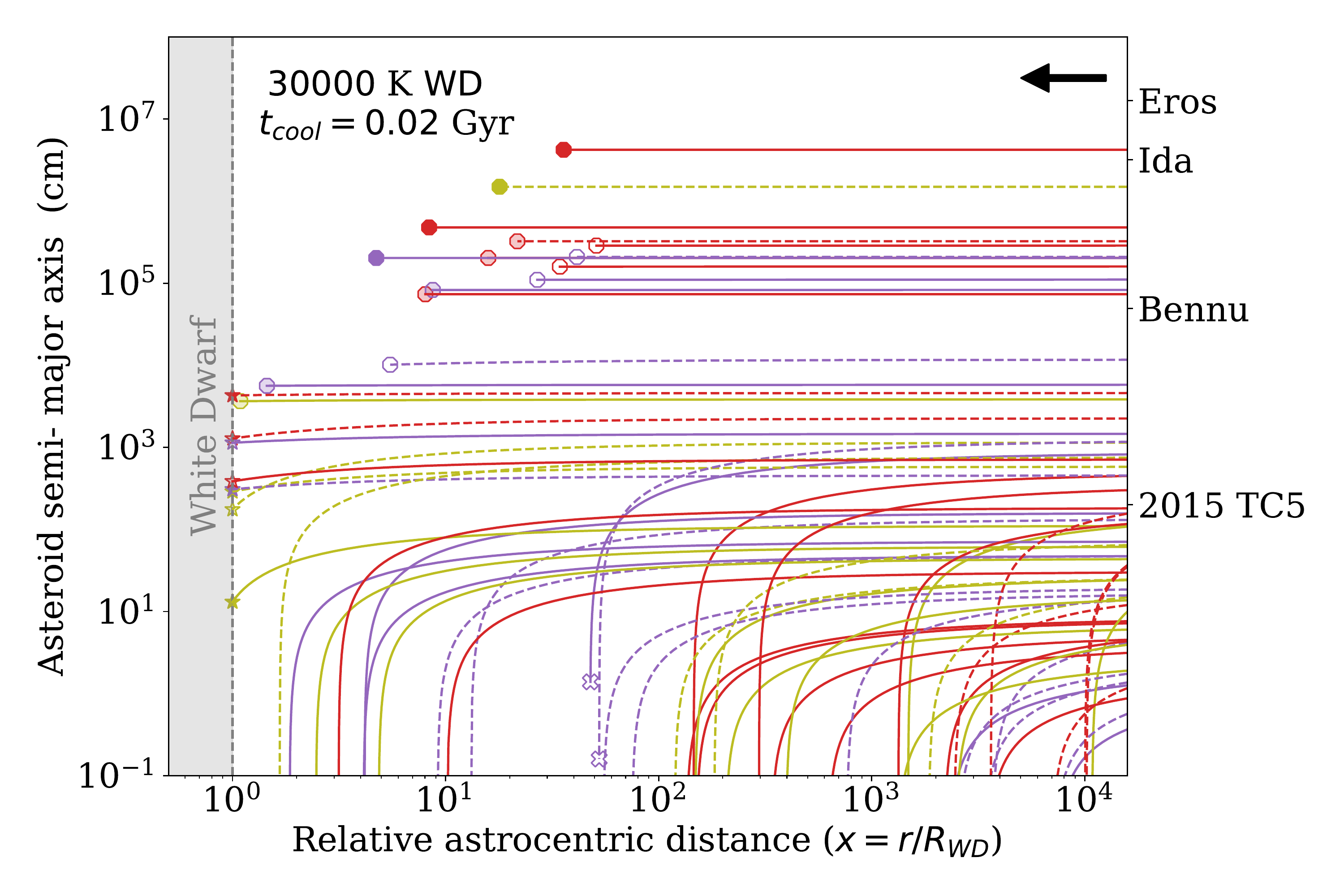}
        \label{subfig:coolWD}
    \end{subfigure}%
    \begin{subfigure}[c]{.5\linewidth}
        \centering
        \includegraphics[scale = 0.299]{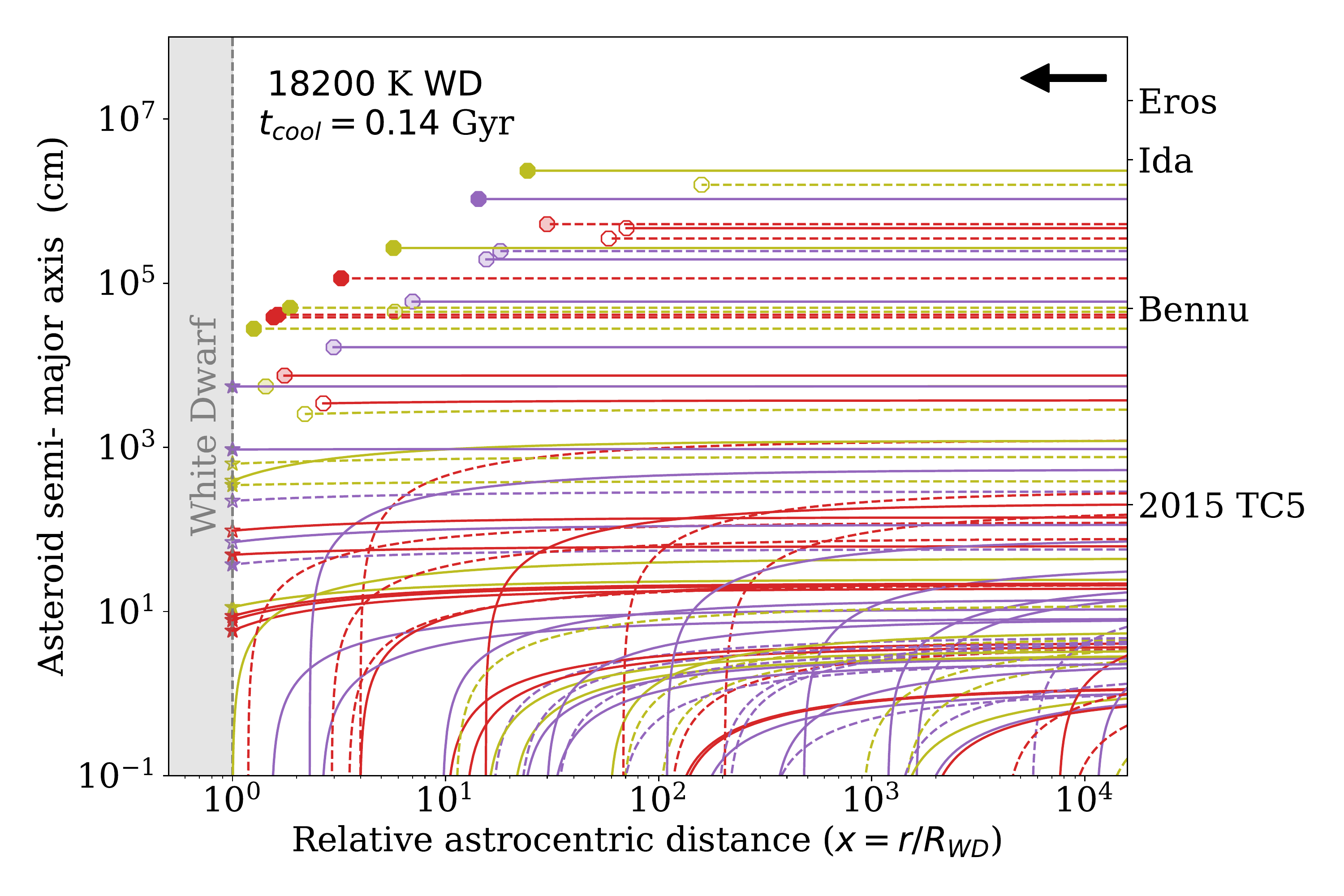}
        \label{subfig:coolishWD}
    \end{subfigure}
    \begin{subfigure}[c]{.5\linewidth}
        \centering
        \includegraphics[scale = 0.299]{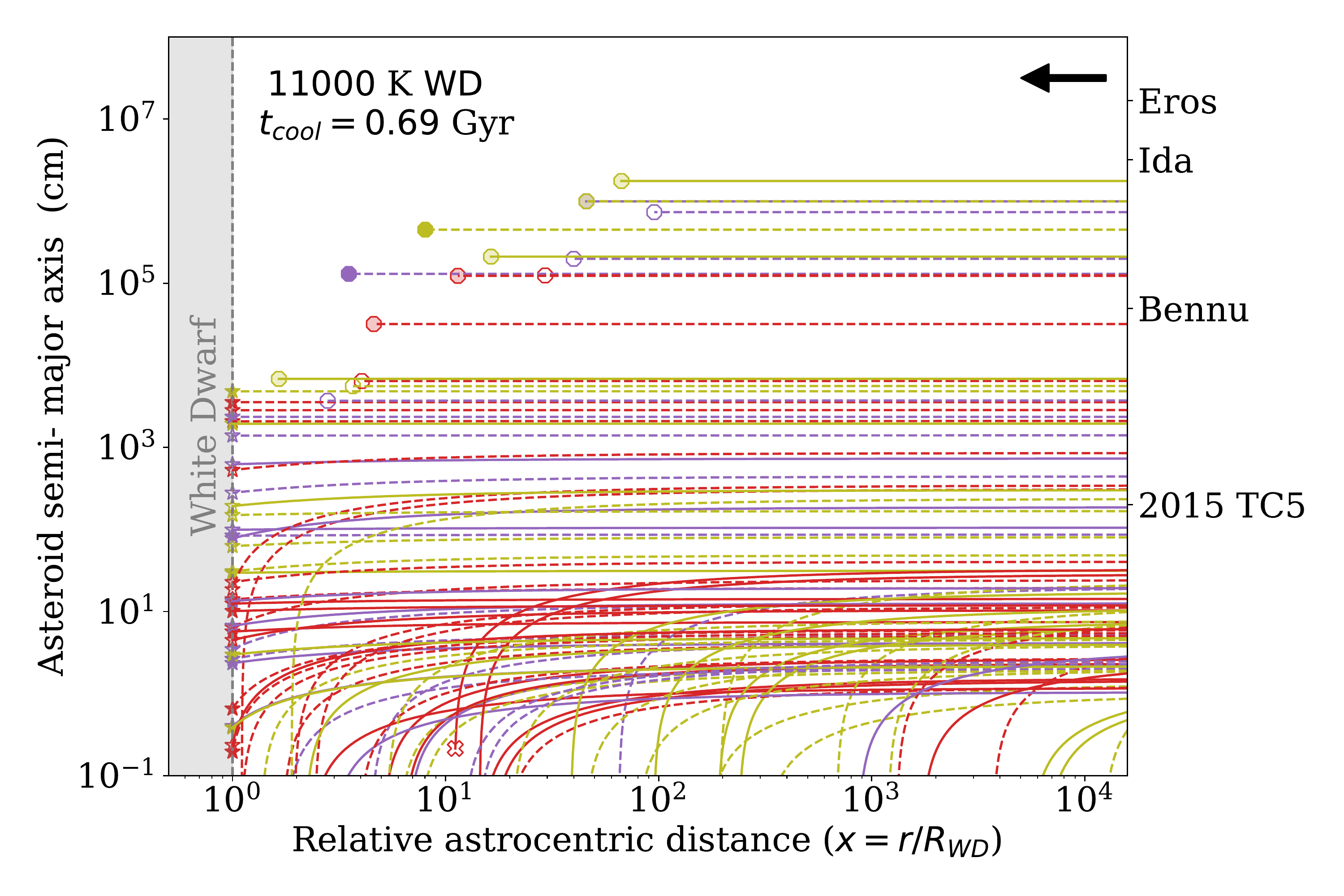}
        \label{subfig:midWD}
    \end{subfigure}%
    \begin{subfigure}[c]{.5\linewidth}
        \centering
        \includegraphics[scale = 0.299]{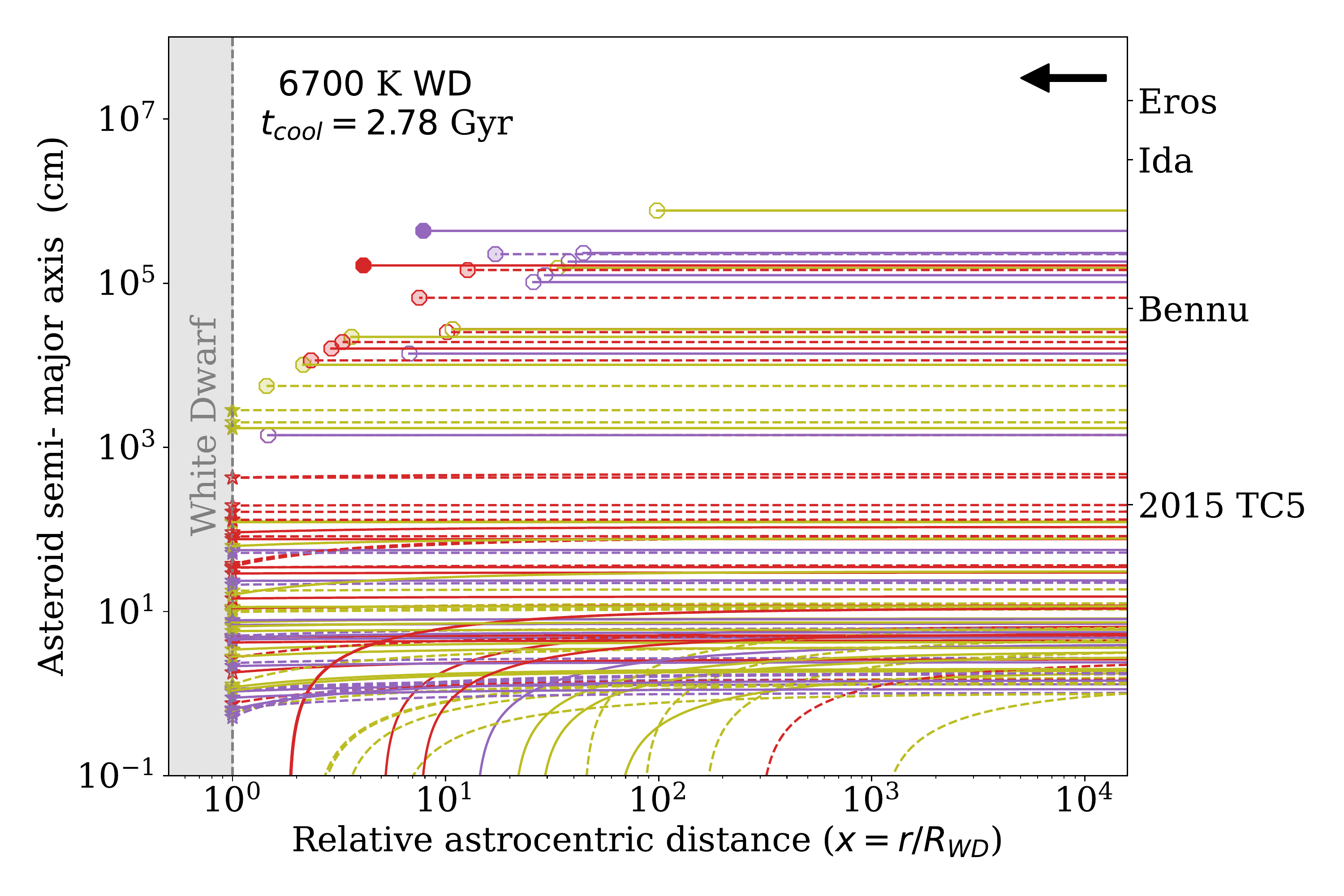}
        \label{subfig:warmWD}
    \end{subfigure}
    \begin{subfigure}[c]{.5\linewidth}
        \centering
        \includegraphics[scale = 0.299]{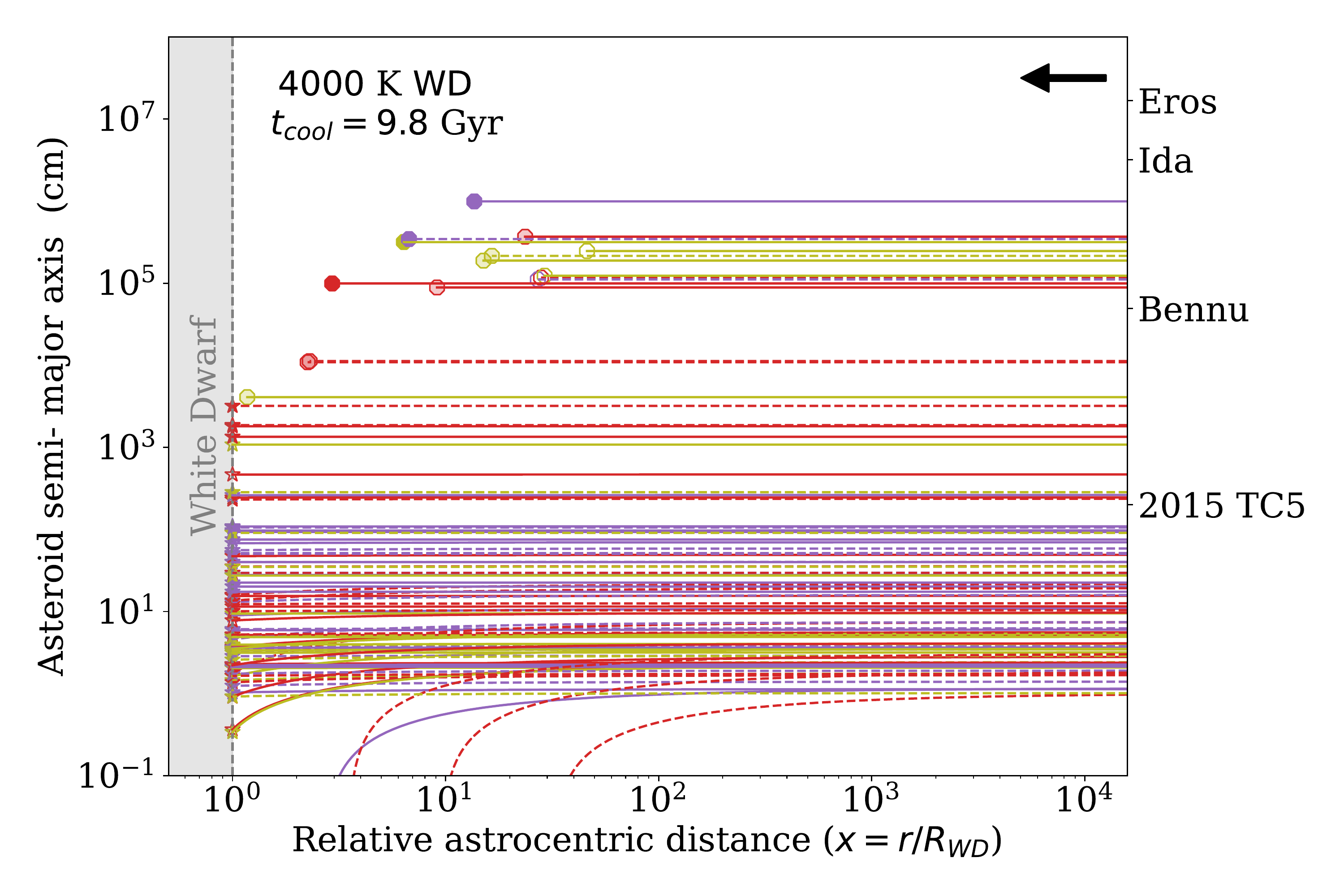}
        \label{subfig:hotWD}
    \end{subfigure}%
    \begin{subfigure}[c]{.5\linewidth}
        \centering
        \includegraphics[scale = 0.3005]{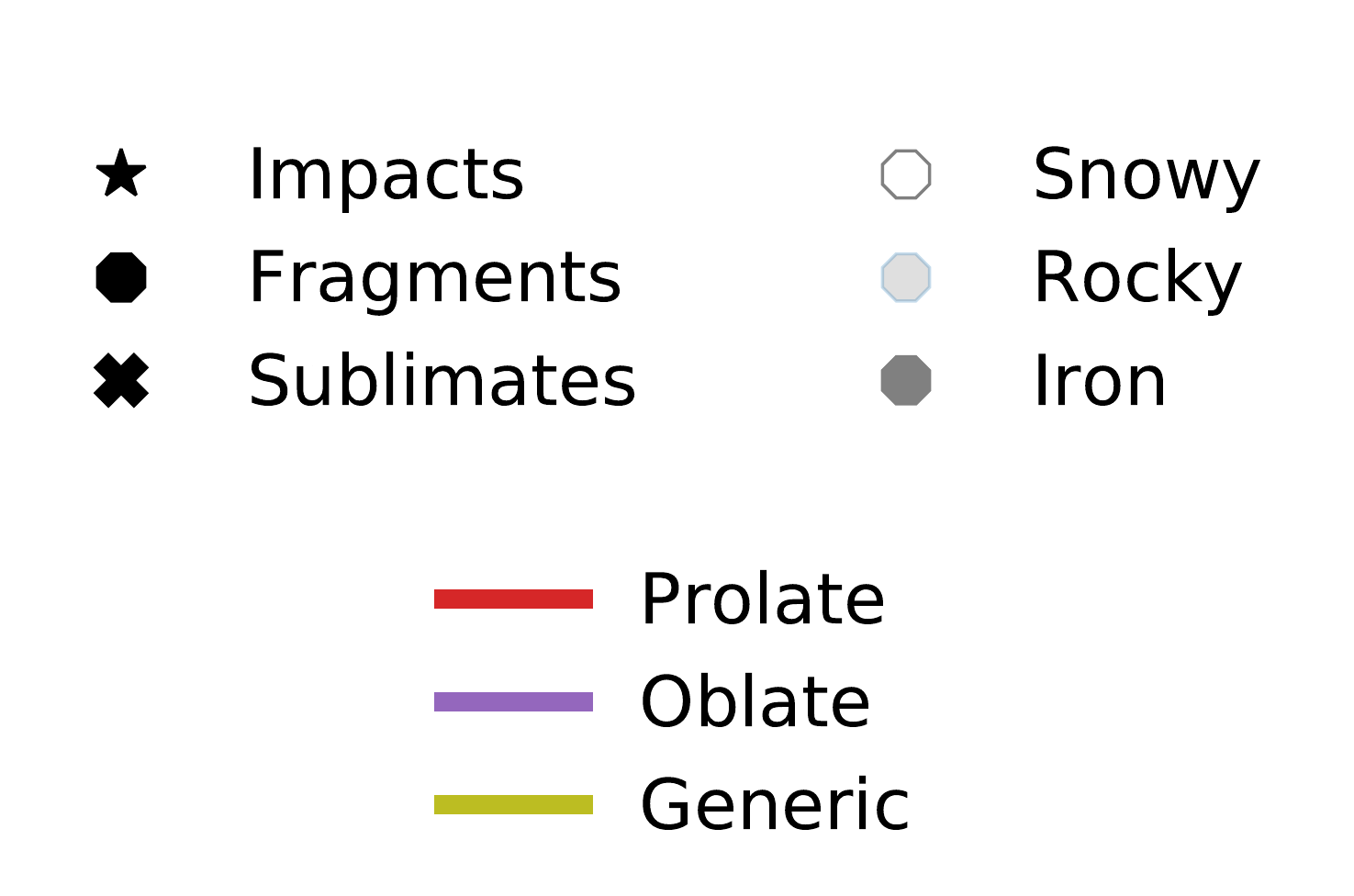}
        \label{subfig:legend}
    \end{subfigure}%
    \caption{The fate of exo-asteroid belts perturbed towards white dwarfs. 
    Each panel shows the entire belt of 100 asteroids with randomly chosen initial largest semi-axis (physical) size, shape and material, for different white dwarf temperatures and cooling ages as stated in the top left hand corner of each panel. 
    The colour of the individual asteroid tracks highlights which shape model has been used, where a solid line indicates the fiducial model, and the dashed line the extreme version.
    The marker at the end of the tracks show the ultimate fate of the body, and the fill of this marker shows the asteroid's material properties.
    The right-hand axis of each plot shows some example sizes of named Solar System asteroids.
    Across all white dwarf temperatures, asteroids above $\sim 10^3$cm in semi-major axis fragment.
    Hotter white dwarfs are able to sublimate larger bodies up to $10^3$cm.
    Cooler white dwarfs are more vulnerable to direct impacts in the size range $10^0-10^4$cm.}
    \label{fig:belts}
\end{figure*}

Fig.~\ref{fig:belts} shows the outcomes for such belts around five white dwarfs with $T_\text{eff} = [30,000$ K, $18,200$ K, $11,000$ K, $6700$ K, $4000$ K] as described in Section~\ref{subsubsec:wdtemp_cool}.
The asteroids enter at the right-hand edge of the plots and move to the left (as indicated by the black arrows), towards the white dwarf which is shown by the grey shaded region at the $y$-axis.
The line colour indicates the shape model from the possibilities detailed in Table~\ref{tab:ellip_shape_models} and Fig.~\ref{fig:shape_models}.
A solid line shows the fiducial shape model has been used, while the dashed line indicates the extreme model. 
The shape marker at the point of disruption illustrates which destruction regime is active:  octagons indicate that the body fragments, crosses are the locations of complete sublimation, and stars at the edge of the white dwarf zone show that the body impacts onto the white dwarf. 
Finally, the fill attribute of these markers demonstrates the material makeup of the bodies: empty is snowy material, transparent is rocky and solid fill is iron material. 
On the right-hand side of each plot some relevant Solar System asteroid sizes are shown\footnote{Sizes for the three largest asteroids are taken from the NASA JPL Small Body Database \url{https://ssd.jpl.nasa.gov/sbdb.cgi}. 2015~TC5 is the smallest named Near-Earth Asteroid \citep{Reddy2016}.}.

Generally it can be seen that regardless of white dwarf temperature and body material, the largest asteroids are likely to fragment across the entire astrocentric distance considered. 
The diagonal lines formed by the fragmentation locations that can be seen in Fig.~\ref{fig:belts} are caused by the fragmentation conditions we impose, and can also be visualized in the coloured lines in Fig.~\ref{fig:crossing_function}.
By recalling that when considering fragmentation we only need to consider the $x$-components of the binding size parameter (equation~\ref{eq:binding_param}), we remove any dependence on the specific shape model used and are left with a relationship for each material, independent of white dwarf temperature.
The weaker the internal strength of the body's material, the further away from the white dwarf the body will fragment. 
The size of the body will also effect the distance of fragmentation, with larger bodies fragmenting further from the central star. 

Although the smallest bodies consistently sublimate completely across all white dwarf temperatures, higher white dwarf temperatures can cause larger bodies to sublimate. 
The two coolest ($4000$ K and $6700$ K) white dwarfs considered here sublimate 4 and 17 per cent of asteroids respectively, this increases to 46 per cent of asteroids sublimated for the $11,000$ K white dwarf. 
The simulations for the two hottest white dwarf temperatures featured in Fig.~\ref{fig:belts} show significantly more sublimation with 61 ($18,200$ K) and 77 ($30,000$ K) per cent of the total belt being completely sublimated. 
This increased level of sublimation also occurs at an increased distance from the white dwarf.
$13$ per cent of the asteroids which sublimate around the $18,200$ K white dwarf do so beyond the maximum relative astrocentric distances displayed in Fig~\ref{fig:belts}.
For the $30,000$ K white dwarf, 32 per cent of total sublimations occur beyond $10^4$ relative astrocentric distances. 

Thus, an additional simulation is carried out for the two hottest white dwarf temperatures with relative astrocentric distance range extended to $10^9$.
\begin{figure*}
    \begin{subfigure}[c]{.5\linewidth}
        \centering
        \includegraphics[scale = 0.3]{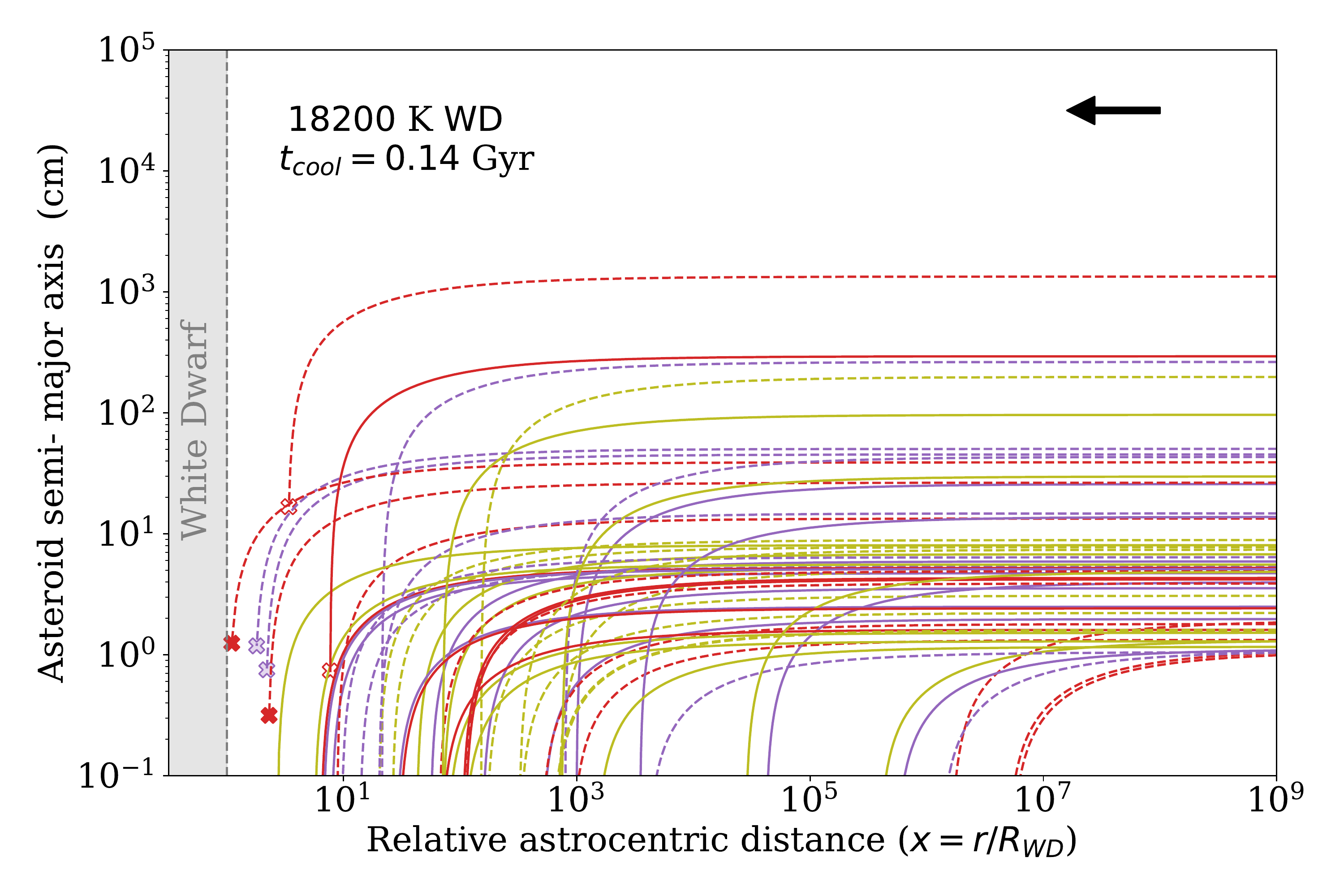}
        \label{subfig:warmWD_extended}
    \end{subfigure}%
    \begin{subfigure}[c]{.5\linewidth}
        \centering
        \includegraphics[scale = 0.3]{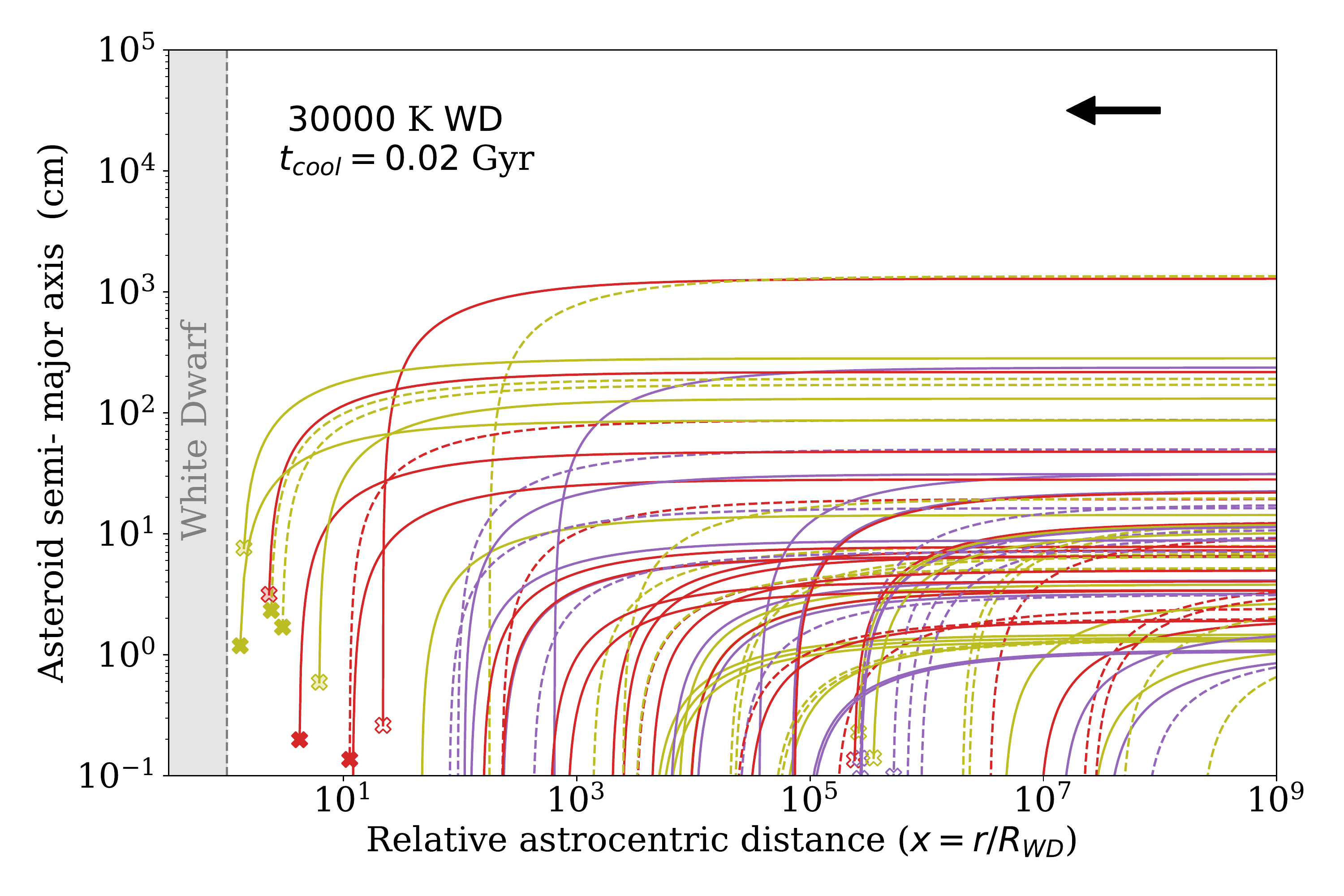}
        \label{subfig:hotWD_extended}
    \end{subfigure}
    \caption{The fate of sublimated Main belt analogue asteroids approaching a white dwarf on an extremely eccentric orbit.
    Here as in Fig.~\ref{fig:belts}, the line colours and line styles indicate the shape model, the marker fill shows the material properties of the specific body. 
    Only sublimated asteroids are shown in these panels, and hence all end points have the same cross-shaped marker.
    The $x$-axis showing the relative astrocentric distance from the white dwarf is extended to $10^9$, as compared to $10^4$ in Fig.~\ref{fig:belts}.
    The hottest white dwarfs considered in this work can completely sublimate small asteroids up to $\sim 10^8$ relative astrocentric distances away from the central star, seven orders of magnitude further out than for the coolest white dwarf considered in this work.}
    \label{fig:extendedbelts}
\end{figure*}
These extended simulations, which can be seen in Fig~\ref{fig:extendedbelts}, show that for hot white dwarfs, small asteroids can sublimate up to $10^8$ relative astrocentric distances from the white dwarf, 4 orders of magnitude larger than for the coolest white dwarf considered here. 
The physical extent at which asteroids can sublimate around a hot white dwarf is particularly interesting when compared to the tidal boundary for canonical rubble pile asteroids. 

In Section~\ref{subsec:fragmentation}, we used the assumption that internal strength dominates over self-gravitation and that bodies are closer to monoliths than rubble piles. 
If we instead adopt the assumption that $\mathcal{\mathbfit{F}_S}/\mathcal{\mathbfit{F}_G}<<1$ then the condition derived by solving equation~(\ref{eq:cond_resist_frag}), provides us with the location of tidal disruption for a rubble pile body,
\begin{equation}
    x_\text{Roche} = \left( \frac{M_\text{WD} a}{\pi \rho U_x R_\text{WD}^3} \right),
    \label{eq:xroche}
\end{equation}
where again we are only considering the $x$-direction.
Using equation~(\ref{eq:xroche}), the material properties given in Table~\ref{tab:props} and the shapes in Table~\ref{tab:ellip_shape_models}, we find that $x_\text{Roche} \sim 10^2-10^4$.
The smaller constituent particles of a rubble pile asteroid are then susceptible to sublimation before they are even tidally disrupted from their parent body, and thus are unlikely to persist for long after an initial disruption event.

Fig.~\ref{fig:belts} also shows that there are very few cases where a body will undergo partial sublimation before impacting directly onto the white dwarf, and no cases where a partially sublimated body fragments.
Two clear examples of partial sublimation can be seen in the top left hand panel of Fig.~\ref{fig:belts} with the two smallest generic shaped asteroids that impact.
These two specific cases begin to lose mass at about 10 relative astrocentric distances from the star and their semi-major axes only decrease by a small amount before they are directly accreted.

For the two hottest white dwarfs considered, the largest body which undergoes partial sublimation is $\sim 10^3$ cm, whereas only the smallest $a < 10^1$ cm bodies partially sublimate around the cooler white dwarfs.
Since our model assumes that the body has a uniform composition, the asteroids here represent a maximal level of sublimation for a particular material, and thus the small amount of objects which partially sublimate could imply a small range of asteroid sizes per white dwarf temperature where volatile elements are preferentially lost compared to more refractory elements.

Although direct impacts occur across all white dwarf temperatures considered, the range of sizes which impact increases as the white dwarf cools.
\begin{table}
	\centering
	\caption{Minimum, median and maximum impactor sizes across the range of white dwarf temperatures.}
	\label{tab:impactors}
	\begin{tabular}{lccr}
		\hline
		 $T_\text{WD}$& Minimum (cm) & Median (cm) & Maximum (cm)\\
		\hline
		$30\,000$ K & 97.9 & 839 & 11400\\
	    $18\,200$ K & 6.72 & 107 & 2290 \\	
	    $11\,000$ K & 0.14 & 12.8 & 10600 \\	  
	    $6\,700$ K & 0.24 & 12.8 & 10600 \\
		$4\,000$ K & 0.84 & 11.7 & 19300\\
		\hline
	\end{tabular}
\end{table}
Table~\ref{tab:impactors} gives the minimum, median and maximum asteroid sizes at moment of impact across all five white dwarf temperatures in the simulations presented in Fig.~\ref{fig:belts}. 
The minimum and median impactor sizes generally decrease as the white dwarf cools. 
This can be explained by the fact that at higher white dwarf temperatures, these smaller bodies will be sublimated completely.
The maximum impact sizes range between the order of $10^3-10^4$ cm, with all of these bodies being made from iron.
The smaller maximum impactor size recorded for the $18,200$ K white dwarf simulation is due to the random nature of size and material selection used in the simulations.
As each simulation has a different selection of size and asteroid properties, the $18,200$ K simulation simply did not contain an iron asteroid with a similar size to the largest impactors in the other simulations.

To confirm the random selection is the cause of the different maximum sizes, a dedicated study with a fixed material and size distribution was carried out. 
The maximum impactor size for iron asteroids is always of the order $10^4$ cm.
For rocky asteroids the maximum is $10^3$ cm, and for snowy asteroids it is $10^2$ cm.
Although it should be noted that for snowy asteroids, no impacts are expected for the hottest white dwarf temperature due to the increased rate of sublimation.
Thus the largest asteroid that could directly impact on a white dwarf's photosphere has $a ~\sim 10^4$ cm = $0.1$ km, a value that is similar in size to the Solar System asteroid Bennu\footnote{Bennu has a radius of $0.246$ km taken from the NASA JPL Small Body Database \url{https://ssd.jpl.nasa.gov/sbdb.cgi}.}.

The maximum asteroid size for direct impact is determined by the specific material properties of the body itself, while the minimum impact size is dictated by the sublimation limit due to the stellar radiation from the white dwarf.
A direct asteroid impact could be inferred from a short term increase in calculated accretion rates. 

\subsection{The effect of shape}
To investigate the role of triaxiality on the destruction regime of a body approaching a white dwarf, we first compare the triaxial model to a spherical model (Section \ref{subsubsec:model_compar}), then we look at the specific impact on sublimation (Section~\ref{subsubsec:shape_sub}).

\subsubsection{Shape model comparison} \label{subsubsec:model_compar}
To further motivate the future use of triaxial shape models, we now directly compare the forces and binding and sublimation parameters introduced in Sections~\ref{subsec:sublimation}-\ref{subsec:fragmentation} to the equivalent spherical forms.

The following spherical forces are similar in form to those presented in BVG17 and \cite{BearSoker2015},
\begin{equation}
    \label{eq:spher_tidal}
    \mathcal{F}_{T_\text{spher}} = \frac{2 G M_\text{WD} M a}{x^3 R_\text{WD}^3},
\end{equation}
\begin{equation}
    \label{eq:spher_strength}
    \mathcal{F}_{S_\text{spher}} = - \pi a^2 S,
\end{equation}
\begin{equation}
    \label{eq:spher_grav}
    \mathcal{F}_{{G}_\text{spher}} = - \frac{G M}{a^2}.
\end{equation}
Using the above forces, we can find the sublimation and binding size parameters (equations \ref{eq:sub_param} and \ref{eq:binding_param}) for a purely spherical shape model
\begin{equation}
    \label{eq:subparam_spher}
    A_\text{spher} = \frac{R_\text{WD}^{3/2} T_\text{eff}^4 \sigma}{2 \mathcal{L} \rho \sqrt{2GM_\text{WD}}},
\end{equation}
\begin{equation}
    \label{eq:bindparam_spher}
    B_\text{spher} = \sqrt{\frac{3 R_\text{WD}^3 S}{8 G M_\text{WD} \rho}},
\end{equation}
where these differ from the BVG17 results only by a numerical factor. 
Remembering that the binding size parameter dictates the fragmentation of an object, that it involves the tidal and strength forces, and that fragmentation always occurs in the $x$-direction of the ellipsoidal body, we can compare the forces in the following ratios
\begin{equation}
    \label{eq:spher_tri_compar_tidal}
    \frac{\mathcal{F}_{T_x}}{\mathcal{F}_{T_\text{spher}}} = 1,
\end{equation}
\begin{equation}
    \label{eq:spher_tri_compar_strength}
    \frac{\mathcal{F}_{S_x}}{\mathcal{F}_{S_\text{spher}}} = \mathfrak{c}\mathfrak{b}. 
\end{equation}

Thus, the tidal force on the longest ($x$) axis of a triaxial body is identical to the spherical body case. 
However, the tensile strength on the longest axis of a triaxial body is reduced by a factor equal to the product of the body's two aspect ratios $\mathfrak{b}$ and $\mathfrak{c}$.
Using these two results and following the procedure from Section~\ref{subsec:fragmentation} we can compare the $x$ component of the triaxial binding size parameter to the spherical parameter
\begin{equation}
    \label{eq:spher_tri_compar_binding}
    \frac{B_x}{B_\text{spher}} = 1.
\end{equation}
The above result confirms that the condition for fragmentation is independent of an individual shape model and can be approximated using a spherical model.

As discussed in Section~\ref{subsec:outcomes}, total sublimation always occurs in the semi-minor axis ($z$-direction) first. 
Using this fact, we can compare $A_z$ from the triaxial model to the spherical sublimation parameter from equation~(\ref{eq:subparam_spher}) to identify the effect of shape on sublimation
\begin{equation}
    \label{eq:spher_tri_compar_sub}
    \frac{A_z}{A_\text{spher}} = \frac{1}{\mathfrak{c}}.
\end{equation}
The result in equation~(\ref{eq:spher_tri_compar_sub}) shows that when using a triaxial shape model, the minimum size a body must be to withstand sublimation all the way to the photosphere of the white dwarf, is increased by a factor of the aspect ratio $\mathfrak{c}$ between the semi-minor and semi-major axes. 
As total sublimation occurs in the semi-minor axes, the $\mathfrak{b}$ aspect ratio between the semi-intermediate and semi-major axes does not affect the level of sublimation in a triaxial asteroid compared to a spherical model.
Thus, triaxial asteroids are more vulnerable to complete sublimation than a spherical asteroid.

\subsubsection{The effect of shape on sublimation} \label{subsubsec:shape_sub}
Figs.~\ref{fig:belts} and \ref{fig:extendedbelts} show that sublimation is most prevalent around hot white dwarfs, where bodies up to $10^4$ cm in size can sublimate completely.
To further investigate the role of triaxiality on sublimation, 1000 asteroids with semi-major axes evenly spaced in log space in the range $10^0-10^4$ cm were run through the analytical process discussed in Section~\ref{subsec:outcomes}, except with fixed shape models and materials for a $18,200$ K and $30,000$ K white dwarf.

\begin{figure*} 
    \centering
    \includegraphics[scale = 0.6]{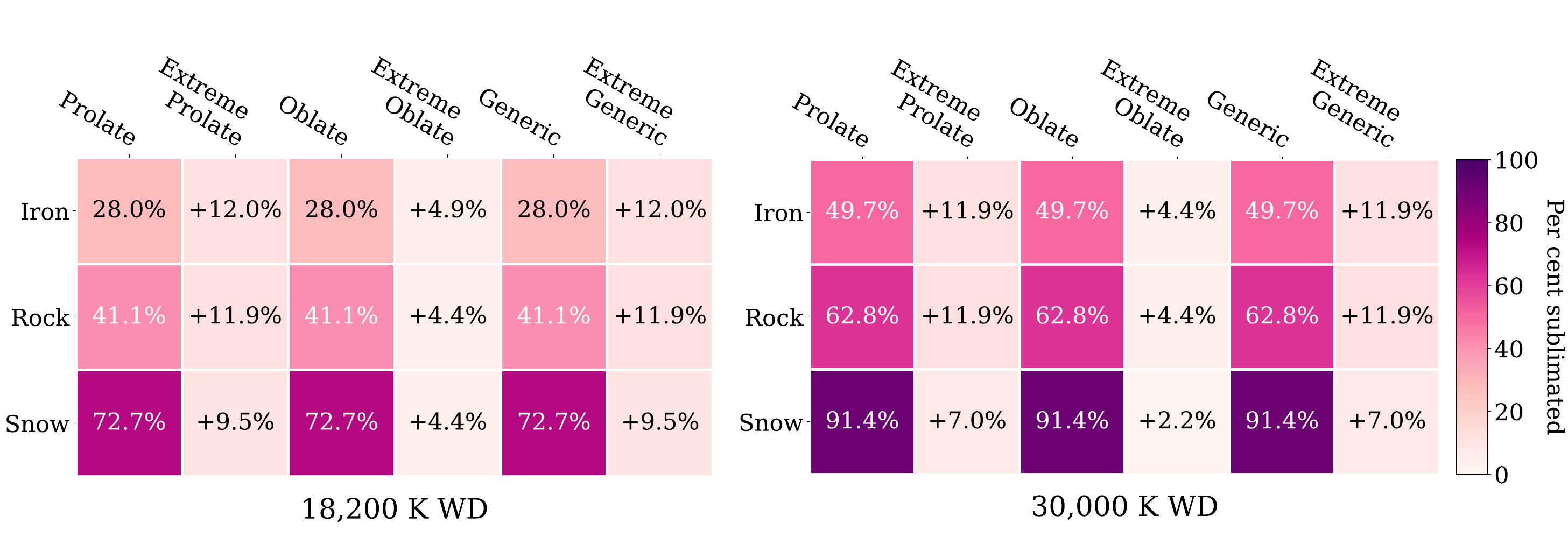}
    \caption{The effect of a triaxial shape model and material on the amount of sublimation an analogue Main belt perturbed towards the white dwarf will undergo.
    The fiducial shape model (prolate, oblate and generic) columns display the percentage of $1000$ asteroids in the size range $10^0-10^4$cm which sublimate completely. 
    The extreme shape model columns show the percentage increase in asteroids which sublimate completely compared to their respective fiducial shape models. 
    The colour of each box illustrates the percentage as described by the colour bar along the right hand side. 
    The fiducial shape models show identical levels of sublimation for each material, all extreme models show even more increased sublimation. 
    The extreme prolate and generic shape models show the same increase in sublimation, although the extreme oblate model exhibits a smaller increase.}
    \label{fig:SubShape}
\end{figure*}

Fig.~\ref{fig:SubShape} shows the fraction of these 1000 asteroids which sublimate for each possible combination of material and shape. 
The fiducial (prolate, oblate and generic) shape columns show the total percentage of bodies which sublimate, while the extreme shape model columns show the percentage increase in sublimation compared to the fiducial models. 
As an example, for the $18,200$ K white dwarf and iron bodies, 28 per cent of prolate bodies sublimated, whereas 40 per cent of extreme prolate asteroids sublimated. 

These graphics show that amongst the fiducial shape models, there is no dependence on shape on the percentage of bodies which sublimate.
However, all of the extreme shape models show an increased level of sublimation compared to their fiducial counterparts. 
The extreme prolate and generic models show the same increase in sublimation, while the extreme oblate models shows less, which can be explained by examining equation~(\ref{eq:spher_tri_compar_sub}) and the shape parameters defined in Table~\ref{tab:ellip_shape_models}.
The fiducial models all share a common aspect ratio $\mathfrak{b} = 0.6$, the extreme oblate model has $\mathfrak{b} = 0.4$, whereas both the extreme generic and extreme prolate models have $\mathfrak{b}= 0.2$. The levels of sublimation increases as the $\mathfrak{b}$ aspect ratio decreases according to equation~(\ref{eq:spher_tri_compar_sub}).

\section{Further considerations} \label{sec:further}
This work takes an overall simple approach to solving the problem of asteroid disruption around a white dwarf. 
The analytical model described here only records the initial disruption process and does not consider what happens to the products of the disruption process. 
The possible subsequent processes which affect the initial disruption products are now briefly considered.

\subsection{Sublimated material} \label{subsec:sublimated}
The sublimated material from an asteroid approaching a white dwarf on an extremely eccentric orbit could quickly accrete onto the white dwarf, or form part of a gaseous debris disc \citep{Trevascus2021}.
The white dwarf SDSS~J1228+1040 is observed with an extremely dense planetesimal orbiting inside a debris disc with a gaseous component expanding out to $\sim 1.2 R_\odot$ \citep{Gansicke2006, Manser2019}.
In order to compare both the level of sublimation and the physical extent of a gas disc which could be produced by the process discussed in this paper, 1000 asteroids are passed through the process described in Section~\ref{subsec:outcomes}.
In this case, we adjust the fiducial white dwarf properties previously used in this paper and adopt the measured SDSS~J1228+1040 properties; $M_\text{WD} = 0.77M_\odot$ and $T_\text{eff} = 22020$ K \citep{Gansicke2006}.
\begin{figure} 
	\includegraphics[width=\columnwidth]{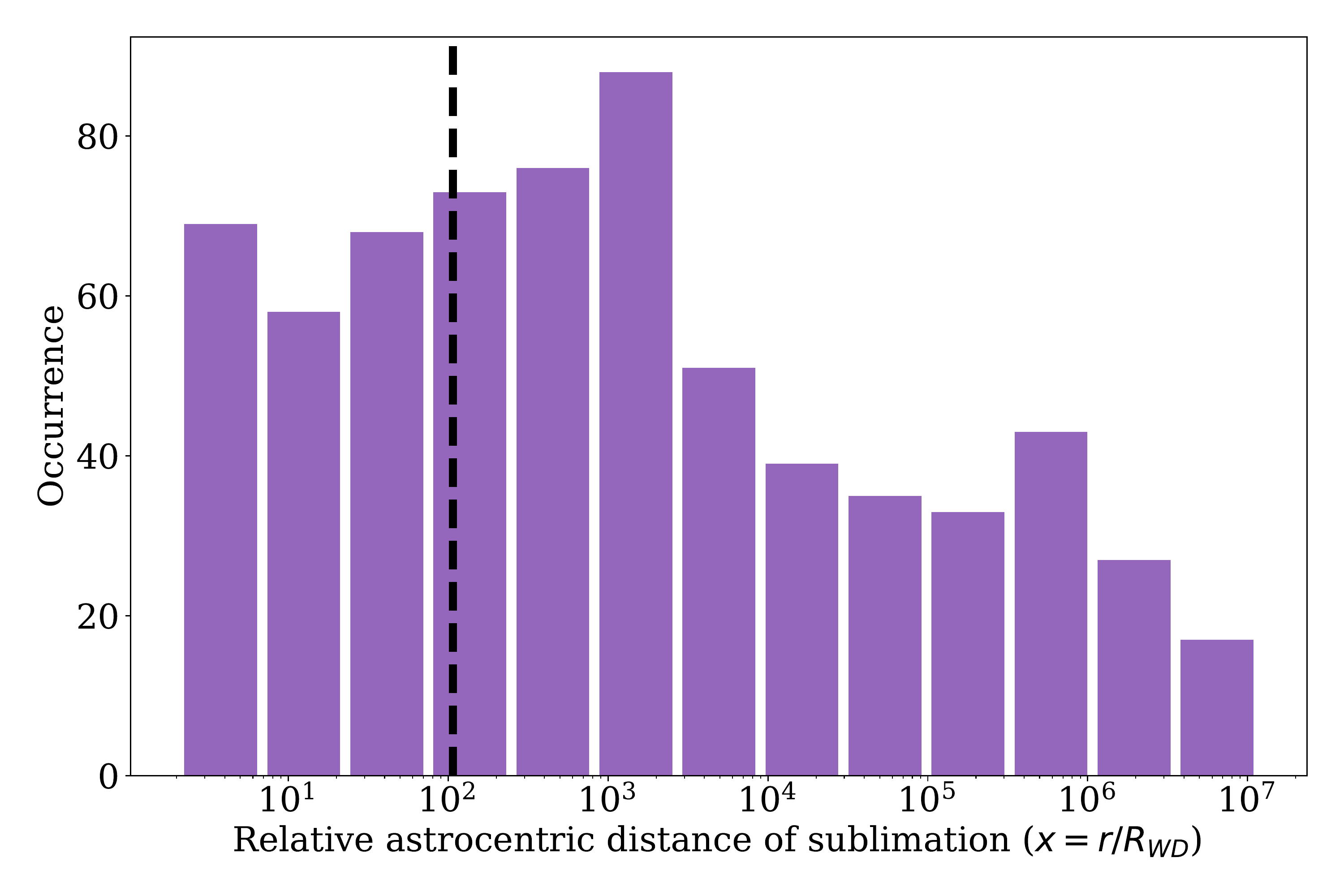}
    \caption{A histogram showing the number of bodies which sublimate at different astrocentric distances from a white dwarf with the same properties as SDSS~J1228+1040. The dashed black line at $\sim 10^2$ indicates the estimated outer radius of the gas disc around SDSS~J1228+1040. This outer radius is similar to those of other gas discs with well-constrained geometries. Small bodies $< 10^4$cm can sublimate far beyond the radial extent of this observed gaseous disc.}
    \label{fig:SDSS1228_compar}
\end{figure}
Fig.~\ref{fig:SDSS1228_compar} shows a histogram of the relative astrocentric distances at which the bodies sublimate. 
The black, dashed, vertical line indicates the outer radius of the SDSS~J1228+1040 gas disc. 
This figure shows that sublimated gaseous materials can easily form within the radial extent of the SDSS~J1228+1040 debris disc. 
Further, gaseous materials could be produced beyond this limit, but with decreasing amounts as the astrocentric distances increases.

This work focusses on sublimation as the origin of gaseous debris around white dwarfs, however this is not the only avenue for gas production. 

Collisional cascades of fragmented material will likely produce gas, both during and immediately after a fragmentation event \citep{Kenyon2017}. 
However, \cite*{Metzger2012} argue that long-lived observations of infrared excesses around white dwarfs preclude collisions being the main source of gaseous debris, as collisions would convert all disc material into gas on the order of days. 
Thus sublimation is expected to play a large role in the production of gaseous debris. 
The interaction of planetesimals with existing gaseous debris can further produce gas as discussed in Section~\ref{subsec:impacted}.

Further, the process of sublimation is complicated if there is a pre-existing debris disc around the white dwarf. 
If an incoming body can be captured and embedded into the debris disc \citep{Grishin2019, OConnor2020, Malamud2021} it can be shielded from sublimative effects.
Since the disc will not be isothermal, the optical depth will vary throughout the disc. 
\cite{Rafikov2012} give the following expressions for the equilibrium temperature of dust particles within an optically thin part of the disc ($T_\text{thin}$) and an optically thick part of the disc ($T_\text{thick}$)
\begin{equation}
    \label{eq:T_thin}
    T_\text{thin} = T_\text{WD} \left( \frac{1}{2} \right)^{1/2} \left( \frac{R_\text{WD}}{r}\right)^{1/2} = T_\text{WD} \left( \frac{1}{2} \right)^{1/2} x^{-1/2},
\end{equation}
\begin{equation}
    \label{eq:T_thick}
    T_\text{thick} = T_\text{WD} \left( \frac{2}{3 \pi} \right)^{1/4} \left( \frac{R_\text{WD}}{r} \right)^{3/4} = T_\text{thick} = T_\text{WD} \left( \frac{2}{3 \pi} \right)^{1/4} x^{-3/4},
\end{equation}
where $x = r/R_\text{WD}$ is the relative astrocentric distance as before. 
$T_\text{thin}$ approximates the temperature of dust particles at the very inner edge of a debris disc which are directly illuminated by the central star.

The canonical model for a white dwarf planetary debris disc is one similar to Saturn's rings; geometrically thin and optically thick.
Thus, $T_\text{thick}$ likely approximates the dust particle temperatures throughout the rest of the disc. 
If we take the outer edge of the SDSS~J1228+1040 gas disc at $R \sim 1.2 R_\odot$ which gives $x \sim 107$ as the outer edge of an optically thick disc, the equilibrium temperature of dust particles around a $T_\text{WD} = 22020$ K white dwarf is $T_\text{thick} \sim 450$ 
K. 
However, the equilibrium temperature in an optically thin disc at the same location is $T_\text{thin} \sim 1505$ K.
Thus, in the presence of a pre-existing debris disc the efficiency of sublimation for infalling bodies is largely dependent on the optical depth of the surrounding material. 
The shielding effects of optically thick material may allow infalling material to come closer to the white dwarf before undergoing sublimation as temperatures rise in the inner disc.

The effect of interactions with an extant debris disc on infalling material which are expected to impact on the white dwarf in this model are further discussed in Section~\ref{subsec:impacted}.

\subsection{Fragmented material} \label{subsec:fragmented}
The products of the fragmentation process will be smaller bodies, whose size will affect what happens to them next.
The products will likely continue fragmenting until the products reach a size where the internal strength of the body will exceed the tidal force acting upon it regardless of how far away it is from the star. 
This resultant distribution of dusty debris will form a ring on short time scales depending on the initial size of the asteroid.
Asteroids which approach the white dwarf from the location of an exo-Main belt at $\approx 5$ au and tidally disrupt, will completely fill a ring with debris within 100 yr \citep{Veras2014a}.

We can provide very rough estimates for the lifetime of a disc of fragments after they have circularized enough \citep{Veras2015b, Nixon2020, Malamud2021} such that their eccentricity and inclination dispersions are less than about 0.1 rad. In this case, if the fragments are assumed to be equal rocky spheres, then we can read off the disc lifetimes from the appropriate figures in \cite{Veras2020d}.

Consider fragmentation at two different locations from a white dwarf: $0.5R_{\odot}$ and $3.0R_{\odot}$, values which are deliberately chosen to straddle the often-used rubble pile Roche limit and to correspond to Figs. 5 and 7 of \cite{Veras2020d}. According to our Fig. \ref{fig:crossing_function}, rocky asteroids with $a \approx 10^{5.8}$ cm and $a \approx 10^{7.0}$ cm will respectively fragment at these distances when $x \approx 40$ and $x \approx 240$, assuming that $R_{\rm WD} \approx 9\times 10^3$ km. These asteroids will form discs of mass $\approx 3 \times 10^{15}$ kg and $\approx 1 \times 10^{19}$ kg, respectively.

Their lifetimes are then determined by the size of the fragments. If $\mathcal{N}$ fragments are formed, then $a_{\rm frag} = \mathcal{N}^{-1/3}a$. Suppose $\mathcal{N} = 10^3$. Then, the disc lifetime for the $a \approx 10^{5.8}$ cm progenitor is $10^{4-5}$ yr. For the $a \approx 10^{7.0}$ cm progenitor, the lifetime depends more strongly on the eccentricity and inclination dispersion of the fragments. When these dispersions are on the order of $10^{-1}$, then the lifetime is $10^{3-5}$ yr. However, for dispersions on the order of $10^{-4}$, the disc lifetime may be comparable to the white dwarf's cooling age.

These disc lifetimes then allow us to provide estimates for the rate material is accreted onto the central white dwarf. 
The less massive $\approx 3\times 10^{15}$~kg debris disc developing from an $a \approx 10^{5.8}$~cm progenitor can thus have accretion rates on the order of $10^5-10^6$~\gs.
However, the larger $a \approx 10^{7.0}$~cm progenitor which leads to a more massive $1\times10^{19}$~kg debris disc can have accretion rates varying from $10^9$~\gs to $10^{11}$~\gs.
Inferred accretion rates for observed white dwarfs lie in the range $10^5 - 10^{10}$~\gs, with older, colder white dwarfs having lower rates \citep[e.g.][]{Wyatt2014, Koester2014, Farihi2016}.
Therefore, our calculated discs can plausibly recreate observed accretion rates.

\subsection{Impactors} \label{subsec:impacted}
As discussed in Section~\ref{subsec:impact} the bodies that enter directly into the white dwarf's photosphere will not be able to survive this encounter and continue on their orbits. 
It is thought that such large-scale accretion could be observed in the form of surface abundance variations for warm white dwarfs. 
DA white dwarfs with $T_\text{eff} > 13,000$K are inefficient at homogenising material accreted onto the surface of the white dwarf and hence one large impact accretion event could be observable in surface abundance variations, whereas DB white dwarfs can homogenise accreted material within a diffusion timescale \citep{Cunningham2021}.
At lower white dwarf temperatures, the homogenisation process becomes more efficient and thus abundance variations from impact events will be unobservable. 

An inherent assumption in our study is that an incoming asteroid does not first impact an extant disc. This important scenario has been considered in several contexts and different regions of parameter space \citep{Grishin2019, OConnor2020, Malamud2021}, and may help to explain the origin of the planetesimal orbiting SDSS~J1228+1040.

Further, the interaction of an incoming asteroid with an existing debris disc can further produce gaseous material \citep{Malamud2021}.
Smaller dust grains entering into a disc can collide with other dust grains resulting in compression shock vaporisation. 
Dust grains colliding with a larger incoming planetesimal can cause dusty material to be ejected from the surface of the body and then go on to vaporise through collisions. 
Energetic gas ions which collide with a planetesimal can set off a cascade of internal collisions which results in atoms being liberated from the surface of the body. 

The gaseous material produced through direct sublimation, as in the focus of this paper, or by the interaction between an incoming body and an extant debris disc as discussed above, can have a further erosive effect on a planetesimal moving within it. 
Bodies which move within a gas disc are subject to a gas drag dependent on the relative velocity of the object and the gas. 
This gas drag can cause outer layers of the body to be lost analagous to aeolian erosive winds \citep{Rozner2021}.

\subsection{Rotation} \label{sec:rotation}
In this work we do not consider the effect of rotation because the tidal potential model used here and in \cite{Dobrovolskis2019} assumes that the minor body is tidally locked to the star and the body's semi-major axis is always pointing towards the central body. 
This assumption has a physical basis because a body's least stable and most vulnerable point to tidal forces is at the end of the longest axis \citep{Harris1996}.
Thus, tidal disruption will always occur in the $x$-direction first, which is seen in Fig~\ref{fig:comparison_plot} and discussed in Section~\ref{subsec:outcomes}.
Since the tidal force on the $x$-axis will always be strongest when pointing directly towards the central body, our results represent the distance furthest from the white dwarf where fragmentation can occur.

Although in this paper we do not consider the effect of asteroid rotation, it is known that if a rubble pile asteroid acquires a sufficient spin rate, it can no longer support itself and undergoes disruption at the so-called `spin barrier' \citep[see fig.~1 of][]{Hestroffer2019}.
While YORP based spin up is expected to destroy a large number of small bodies during the giant branch phases, it has also been shown that extremely eccentric asteroids can chaotically increase their rotational speed through the exchange of orbital and angular momentum at repeated pericentre passages \citep{Makarov2019, Veras2020a}. 
Rotation thus provides a whole new avenue to destruction for eccentric, triaxial asteroids that is not considered here. 

\section{Conclusions} \label{sec:conclusions}
Increasing observations of minor bodies being disrupted around white dwarfs provide motivation for increasing our understanding of the processes which lead to these bodies being destroyed.
Most previous theoretical work on this topic has used spherical shape models to approximate asteroids. However, Solar System studies show that asteroids can be well-approximated by triaxial ellipsoids.
In this work, we expand on the work of \cite{Brown2017} studying steeply infalling debris around a white dwarf by considering the effects of a triaxial shape model on the destruction mode of an asteroid approaching a white dwarf on an extremely eccentric orbit using analytical methods and considering an ensemble of asteroids from simplified Main belt analogues. 
To consider the effect of a white dwarf's temperature on the type of disruption, we first provide an empirical relation between the white dwarf cooling age and effective temperature which encompasses both DA and DB cooling models in equation~\ref{eq:cooling_age_approx}. 

By considering the individual forces acting on each individual principal direction of the body, we define the binding size parameter (equation~\ref{eq:binding_param}) as the size a body must exceed to fragment and the sublimation parameter (equation~\ref{eq:sub_param}) as the minimum size a body must be to survive sublimation to the white dwarf photosphere.
These two parameters allow us to outline an analytical framework to quickly identify how and where an asteroid with specific properties will disrupt around a white dwarf, this framework is shown graphically in Fig.~\ref{fig:flowchart}. 

Using this analytical model, we identified that tidal fragmentation principally occurs in the largest semi-axis for bodies larger than $\sim 100$m (Fig.~\ref{fig:comparison_plot}).
As considering the semi-major axis, $x$-direction, for a triaxial shape model is equivalent to considering a spherical model with radius the same size as the semi-major axis, using a spherical shape model is adequate to investigate tidal disruption. 

On the other hand, total sublimation will occur first in the smallest semi-axis (Fig.~\ref{fig:comparison_plot}), and thus in order to not underestimate the distance from the white dwarf where sublimation occurs, an ellipsoidal shape model should be used.
Hot white dwarfs can sublimate bodies up to $\sim 10^3$cm at large distances from the white dwarf (Fig.~\ref{fig:extendedbelts}), beyond the estimated extent of the gaseous debris disc around the white dwarf SDSS~J1228+1040 (Fig.~\ref{fig:SDSS1228_compar}). 
Cooler white dwarfs are only efficient at sublimating extremely small bodies ($a < 10^1$cm) at distances relatively close to the white dwarf (Fig.~\ref{fig:belts}). 
Snowy, cometary, bodies are more susceptible to sublimation than rock or iron bodies which have higher values of latent heat. 
The fiducial triaxial shape models used in this work have little effect on the amount of sublimation. 
However, the extreme shape models with a greater degree of elongation show increased levels of sublimation (Fig.~\ref{fig:SubShape}).
The increased level of sublimation is caused by the minimum size a body must be to withstand sublimation increasing by a factor of $\mathfrak{c}$, the ratio between the longest and shortest semi-axes, for triaxial models compared to spherical as in equation~\ref{eq:spher_tri_compar_sub}.

Bodies which neither fragment nor sublimate can directly impact the white dwarf's photosphere. 
The minimum impactor size is governed by the maximum size body that the white dwarf can sublimate, and hence the temperature of the white dwarf. 
The maximum impactor size depends on the minimum body size which will fragment while approaching the white dwarf as seen in Table~\ref{tab:impactors}. 
Thus the maximum impactor size is independent of the white dwarf temperature and the minimum size increases as the white dwarf cools. 

To investigate how the planetary debris around a white dwarf would change as the star cools and ages, we simulated simplified Main belt analogues of 100 bodies with sizes drawn randomly between $10^0 - 10^9$cm and assumed all the bodies were randomly perturbed towards the white dwarf without further dynamical interactions (Fig.~\ref{fig:belts}). 
The material properties of the bodies were chosen to broadly align with three different planetary materials; snowy-cometary bodies, rocky bodies similar to meteorites and solid iron bodies and largely affect the destruction outcomes.

It was found that early in a white dwarf's lifetime, while it still has a relatively large effective temperature, bodies of 10s of metres can sublimate completely at distances quite far from the white dwarf.  
The condition for a body to fragment is affected by the size of the white dwarf, but not by its temperature (equation~\ref{eq:binding_param}). 
Therefore, across all ages, bodies larger than $\sim 100$m can fragment. 
The bodies that survive either of these conditions will enter directly into the white dwarf's photosphere.

Ultimately, the cooling age, and hence effective temperature, of the white dwarf can have a large effect on the distribution of any disrupted material. 
While white dwarfs are young and hot, a broader ring of gaseous sublimated material out to large distances ($10^9$ relative astrocentric distances) could be expected (Fig.~\ref{fig:extendedbelts}).
The physical extent of solid, tidal fragments would not be different between white dwarf cooling ages. 
The size range of bodies which directly impact onto the white dwarf grows as the white dwarf cools, however, the possibility of observing such direct impacts from variations in surface abundances decreases as the white dwarf cools. 

\section*{Acknowledgements}
We thank the reviewer, Ed Young, for thoughtful comments which have greatly improved the manuscript.
DV gratefully acknowledges the support of the STFC via an Ernest Rutherford Fellowship (grant ST/P003850/1). We thank Tim Cunningham and Christopher J. Manser for helpful and insightful discussions.

\section*{Data Availability}
The simulation inputs and results discussed in this paper are available upon reasonable request to the corresponding author. 




\bibliographystyle{mnras}
\bibliography{TriaxialAsteroids} 



\appendix

\section{The Sublimation process} \label{app:sub}
Here we present the full derivation for the variation in the asteroid's semi-major axis size due to the effects of sublimation (equation~\ref{eq:a_sub}), to aid the reader in understanding our sublimation model.

The radiation flux at an astrocentric distance $r$ (with $x = r/R_*$ as in equation~\ref{eq:astero_dist}) is
\begin{equation}
    \label{eq:radflux}
    F_\text{rad}\left(r\right) = \frac{L_\text{WD}}{4 \pi r^2} = \left[\frac{R_\text{WD}}{r}\right]^2 \sigma T_\text{eff}^4 = \frac{F_\text{WD}}{x^2},
\end{equation}
where $L_\text{WD}$ is the bolometric luminosity of the white dwarf,  $R_\text{WD}$ is the radius of the white dwarf, $T_\text{eff}$is the white dwarf effective temperature, $F_\text{WD}$ is the bolometric radiation flux at the surface of the star and $\sigma$ is the Stefan-Boltzmann constant. 
The power of incident starlight given by equation~\ref{eq:incident_power} is simply $F_\text{rad} \times \text{area}$.

A simple expression for the mass loss per unit radial distance, which assumes that sublimation occurs at its maximum rate and does not take into account the intrinsic vapour pressure, interactions with an extant accretion disk or other effects which might alter the sublimation process, can be found as follows
\begin{equation}
    \begin{aligned}
    \label{eq:massloss_radialdist}
        \frac{\mathrm{d}\mathbfit{M}}{\mathrm{d}r} &= \frac{1}{v(r)} \frac{\mathrm{d}\mathbfit{M}}{\mathrm{d}t} = \frac{1}{v(r)} \frac{\mathbfit{P}_*}{\mathcal{L}}  \\ &= \frac{1}{v(r)}\frac{a^2 \pi T_\text{eff}^4 \sigma }{\mathcal{L} x^2} \left[ \mathfrak{b} \mathfrak{c} \hat{i} + \mathfrak{c} \hat{j} + \mathfrak{b} \hat{k} \right].
    \end{aligned}
\end{equation}

Using the definitions of the orbital velocity (equation~\ref{eq:orbital_velocity}) and the ellipsoidal mass (equation~\ref{eq:asteroid_mass}) to rewrite equation~\ref{eq:massloss_radialdist}, we can find the mass loss per astrocentric distance ($x = r/R_*$ equation~\ref{eq:astero_dist}) as follows
\begin{equation}
    \begin{aligned}
    \label{eq:masschange_astrodist}
    \frac{\mathrm{d}\mathbfit{M}}{\mathrm{d}x} &= \frac{\mathrm{d}\mathbfit{M}}{\mathrm{d}r} \frac{\mathrm{d}r}{\mathrm{d}x} \\ 
    &= \frac{\pi \sigma T_\text{eff}^4 R_\text{WD}}{\mathcal{L} v_* x^{3/2}} a^2  \left[ \mathfrak{b} \mathfrak{c} \hat{i} + \mathfrak{c} \hat{j} + \mathfrak{b} \hat{k} \right].
    \end{aligned}
\end{equation}

Finally we can write the change in largest semi-axis $a$ per astrocentric distance $x$ due to sublimative forces on the three principal axes,
\begin{equation}
    \begin{aligned}
        \label{eq:achange_astrodist}
        \frac{\mathrm{d}a}{\mathrm{d}x} &=
        \frac{\mathrm{d}a}{\mathrm{d}\mathbfit{M}}\frac{\mathrm{d}\mathbfit{M}}{\mathrm{d}x} \\ 
        &= \frac{R_\text{WD}^{3/2} T_\text{eff}^4 \sigma}{2^{5/2} (GM_\text{WD})^{1/2} \mathcal{L} \rho x^{3/2}} \left[ \hat{i} + \frac{1}{\mathfrak{b}} \hat{j} + \frac{1}{\mathfrak{c}}\hat{k} \right].
    \end{aligned}
\end{equation}
This equation allows us to find the largest semi-axis $a$ as a function of astrocentric distance $a_\text{sub}(x)$ as in equation~\ref{eq:a_sub}.

\section{The $\alpha$-$\beta$ Plane} \label{app:outcomes}
Here we outline a process which can be followed to quickly identify how and where a particular asteroid may disrupt, which may be of use to others.
In Section~\ref{subsec:outcomes} we follow the size evolution of an asteroid approaching the white dwarf across a fine grid of relative astrocentric distances.
However, it is also possible to identify where and how a particular asteroid will undergo destruction using only the sublimation (equation~\ref{eq:sub_param}) and binding size (equation~\ref{eq:binding_param}) parameters.
Both of these parameters can be converted into a dimensionless form by dividing by the body's initial largest semi-axis $a_0$
\begin{equation}
    \label{eq:dimenless_A}
    \boldsymbol{\alpha} = \frac{\mathbfit{A}}{a_0},
\end{equation}
\begin{equation}
    \label{eq:dimenless_B}
    \boldsymbol{\beta} = \frac{\mathbfit{B}}{a_0}.
\end{equation}
These dimensionless quantities allow us to further examine the conditions for fragmentation, sublimation and impact.
If we remember that the condition for fragmentation to occur is that the fragmentation and sublimation sizes are equal $a_\text{frag}(x) = a_\text{sub}(x)$, we can write the intersection of the two functions as
\begin{equation}
    \label{eq:crossing_func}
    f_\text{cross}(x) = \frac{A_x}{x^{1/2}} + B_x x^{3/2} = a_0.
\end{equation}
This function is U-shaped, with a minimum, $a_\text{crit}$, that occurs when $f'_\text{cross}(x) = 0$ at the following points
\begin{equation}
    \label{eq:xcrit}
    x_\text{crit} = \left(\frac{A_x}{3B_x} \right)^{1/2}
\end{equation}
and
\begin{equation}
    \label{eq:acrit}
    a_\text{crit} = \Gamma A_x^{3/4} B_x^{1/4} = \Gamma a_0 \alpha_x^{3/4} \beta_x^{1/4},
\end{equation}
where
\begin{equation}
    \Gamma = \left[ 3^{1/4} + 3^{-3/4} \right] \simeq 1.75.
\end{equation}
There are two possible solutions to equation~(\ref{eq:crossing_func}), with the larger solution, $x_2$, representing the location of the onset of fragmentation, since the asteroid reaches that point before $x_1$.
Thus, the first condition that must be met for fragmentation to occur is $a_0 > a_\text{crit}$.

\begin{figure}
	\includegraphics[width=\columnwidth]{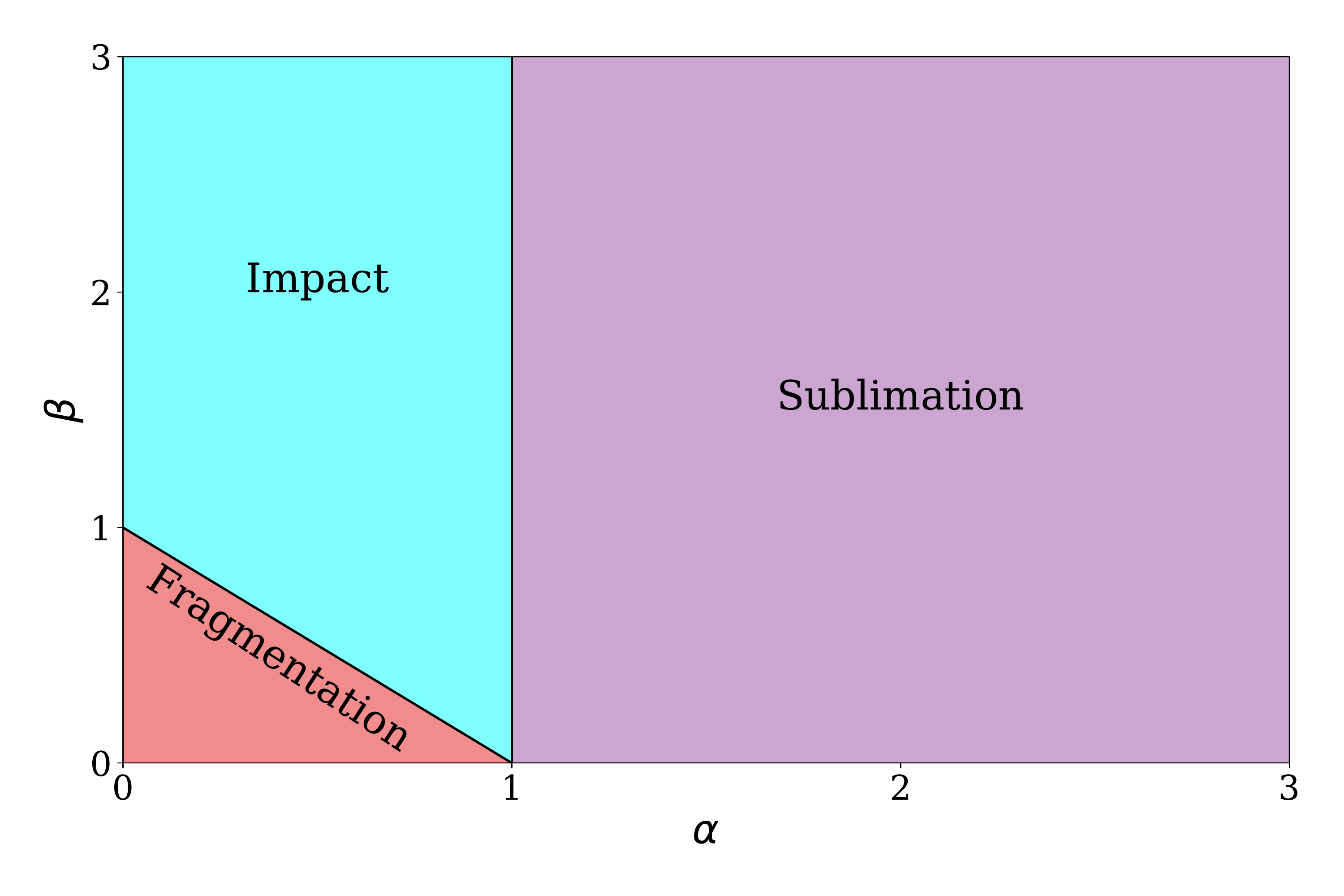}
    \caption{The possible destruction outcomes in the $\alpha-\beta$ plane: total sublimation, fragmentation and direct impact.
    Fragmentation is restricted to the lower left hand corner of the phase space where both $\alpha$ and $\beta$ are less than 1. 
    Sublimation occurs whenever $\alpha$ is larger than one and impact occurs when neither of these two conditions are met.}
    \label{fig:phase_space}
\end{figure}

The second, more stringent, fragmentation condition is that fragmentation occurs outside of the white dwarf photosphere, with $x_2 > 1$.
Thus fragmentation can only occur if both of the following conditions are satisfied
\begin{equation}
    \label{eq:frag_condition}
    \begin{aligned}
     A_x + B_x &< a_0, \\
    \alpha_x + \beta_x &< 1.
    \end{aligned}
\end{equation}
The remainder of the $\alpha$-$\beta$ domain is simply divided into impact or sublimation along the line $\alpha = 1$, where the objects with $\alpha < 1$ can survive sublimation. 
The individual destruction regimes in $\alpha$-$\beta$ space are shown in Fig.~\ref{fig:phase_space}.

\begin{figure} 
	\includegraphics[width=\columnwidth]{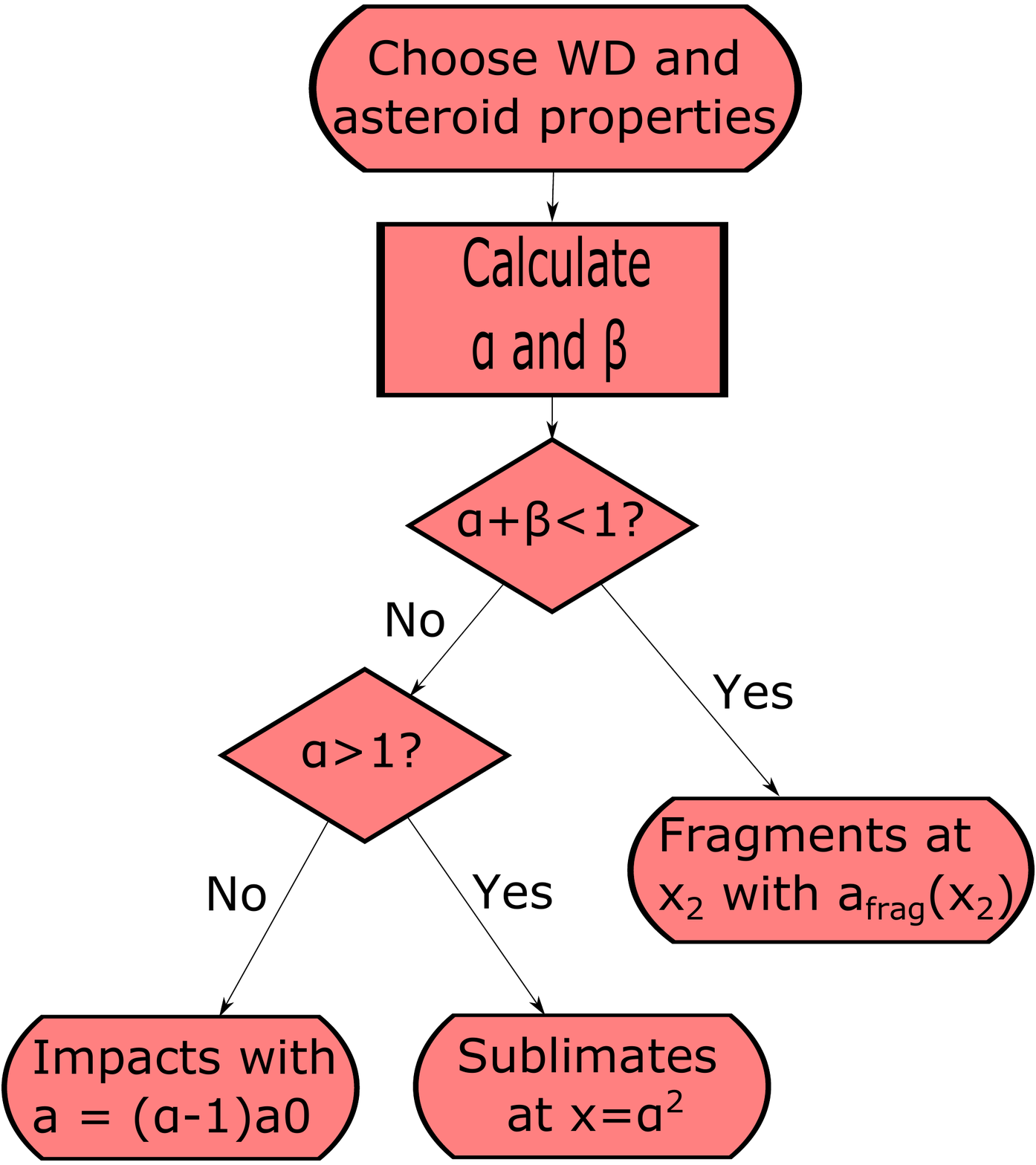}
    \caption{A flowchart which shows how to find the destruction regime, size and position of the failure for any arbitrary selection of white dwarf and asteroid properties. The shape of the asteroid is embedded within the values of $\alpha$ and $\beta$.}
    \label{fig:flowchart}
\end{figure}

Although the further analysis in this paper will track the asteroid's size across a fine grid of astrocentric values, there is an alternative method to determine which destruction regime is relevant.
Again, this alternative method comes down to finding $\alpha$ and $\beta$ for any combination of white dwarf and asteroid properties.
Once these two values are known, a logical process as described in Fig.~\ref{fig:flowchart} can be carried out in each principal direction to identify which form of disruption occurs.
Whichever principal axis disrupts at the largest relative astrocentric distance, $x$, will be the ultimate disruption mode.
If the outcome is fragmentation, the position $x_2$ can be found from a look-up resource.
\begin{figure} 
	\includegraphics[width=\columnwidth]{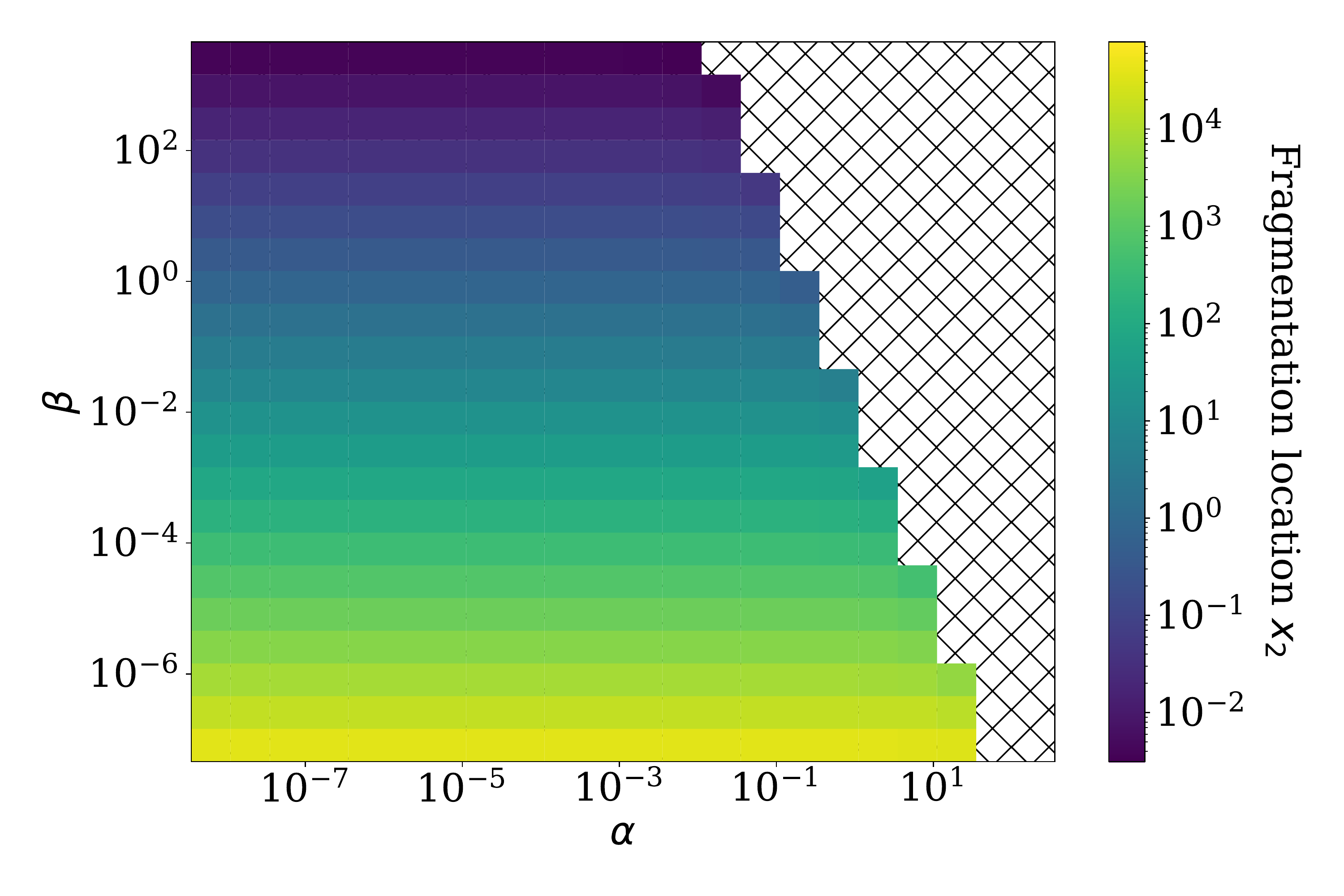}
    \caption{$x_2$ values for a range of $\alpha$ and $\beta$ values. The $x_2$ values are indicated by the colour, which is described in the colour bar on the right hand side of the plot.
    The hatched area with the white background indicates that there is no fragmentation solution for that particular pair of $\alpha$ and $\beta$ values. 
    Smaller values of both $\alpha$ and $\beta$ trigger the fragmentation of the asteroid at greater $x_2$ values further from the white dwarf.}
    \label{fig:ab_pairs}
\end{figure}
Such a resource could either take the form of a table of values such as presented in BVG17, or a plot of different $x_2$ values for pairs $\alpha$ and $\beta$ values as can be seen in Fig~\ref{fig:ab_pairs}.


\bsp	
\label{lastpage}
\end{document}